\documentclass[twocolumn]{aastex631}
\usepackage{amsmath}
\usepackage{amssymb}
\usepackage{bm}
\usepackage{booktabs}
\usepackage{multirow}
\usepackage{hyperref}
\shortauthors{Park et al.}

\graphicspath{{./}{figures/}}

\begin{document}

\title{Quantification of high dimensional non--Gaussianities and\\ its implication to Fisher analysis in cosmology \footnote{Released on \today}}

\correspondingauthor{Core Francisco Park}
\email{corefranciscopark@g.harvard.edu}

\author[0000-0002-9542-2913]{Core Francisco Park}
\affiliation{Harvard University, 17 Oxford St., Cambridge, MA 02138, USA}

\author[0000-0003-3755-7593]{Erwan Allys}
\affiliation{Laboratoire de Physique de l'École Normale Supérieure, ENS, Université PSL, CNRS, Sorbonne Université, \\ 
Université Paris Cité, 75005 Paris, France}

\author[0000-0002-4816-0455]{Francisco Villaescusa-Navarro}
\affiliation{Center for Computational Astrophysics, Flatiron Institute, 162 5th Avenue, New York, NY, 10010, USA}
\affiliation{Department of Astrophysical Sciences, Princeton University, Peyton Hall, Princeton NJ 08544, USA}

\author[0000-0003-2808-275X]{Douglas Finkbeiner}
\affiliation{Harvard University, 17 Oxford St., Cambridge, MA 02138, USA}

\begin{abstract}

It is well known that the power spectrum is not able to fully characterize the statistical properties of non-Gaussian density fields. Recently, many different statistics have been proposed to extract information from non-Gaussian cosmological fields that perform better than the power spectrum. The Fisher matrix formalism is commonly used to quantify the accuracy with which a given statistic can constrain the value of the cosmological parameters. However, these calculations typically rely on the assumption that the likelihood of the considered statistic follows a multivariate Gaussian distribution. In this work we follow \cite{sellentin_2017} and use two different statistical tests to identify non-Gaussianities in different statistics such as the power spectrum, bispectrum, marked power spectrum, and wavelet scatering transform (WST). We remove the non-Gaussian components of the different statistics and perform Fisher matrix calculations with the \textit{Gaussianized} statistics using Quijote simulations. We show that constraints on the parameters can change by a factor of $\sim 2$ in some cases. We show with simple examples how statistics that do not follow a multivariate Gaussian distribution can achieve artificially tight bounds on the cosmological parameters when using the Fisher matrix formalism. We think that the non-Gaussian tests used in this work represent a powerful tool to quantify the robustness of Fisher matrix calculations and their underlying assumptions. We release the code used to compute the power spectra, bispectra, and WST that can be run on both CPUs and GPUs.

\end{abstract}


\section{Introduction} \label{sec:intro}

Upcoming surveys of the large scale structure (LSS) of the Universe like DESI \citep{levi2013desi}, Euclid \citep{laureijs2011euclid}, Rubin Observatory \citep{jain2015greater,lsstsciencecollaboration2009lsst,lsstdarkenergysciencecollaboration2012large} will map the distribution of galaxies in angular and redshift space over large cosmological volumes. 
These galaxies will serve as a biased tracer of the underlying matter density field. If this field were an homogeneous Gaussian random field, the power spectrum would contain all the information about the cosmological parameters. 
However, the matter density field today or at low redshift is highly non Gaussian, especially at the small scales ($\lesssim10\:h^{-1}\rm{Mpc}$) and the power spectrum is not able to fully characterize the statistical properties of it. 

Recently, different methods and statistics have been developed to efficiently extract the cosmological information hidden in the matter, halo, and galaxy density fields \citep{Villaescusa_Navarro_2020, Samushia_2021, Gualdi_2021, Kuruvilla_2021, Bayer_2021,  Banerjee_2019, Hahn_2020, Uhlemann_2020, Friedrich_2020, massara_2021, Dai_2020, Allys_2020, Banerjee_2020, Banerjee_2021, Gualdi_2020, Gualdi_2021, Giri_2020, Bella_2020, Hahn_2021, Valgiannis_2021, Bayer_2021, Kuruvilla_2021b, Naidoo_2021, Porth_2021, Harnois_2021, Liu_2019, Zack_2019, Will_2019, Marques_2019, Lee_2020a, Lee_2020, Gemma_2019, Ajani_2020, Harnois_2021a, Sihao_2021, Harnois_2021, vicinanaza_2019, simpson_2011, simpson_2013, neyrinck_2009, Liu_2022}. For instance, \cite{Hahn_2020} uses the halo bispectrum to break the parameter degeneracy between $\sigma_8$ and $M_{\nu}$ and shows that the sum of neutrino masses can be measured with $\sim5\times$ higher accuracy than just using the power spectrum.
\cite{vicinanaza_2019} evaluates the Minkowski functionals of lensing convergence maps which are helpful breaking the $\Omega_m$-$\sigma_8$ degeneracy. 
Other promising approaches consists of applying a simple non--linear input transform to the density field. 
\cite{simpson_2011,simpson_2013} clips the density field to a maximum value to reduce the large contribution of massive halos to the power spectrum, while \cite{neyrinck_2009} log transforms the density field to weight all elements of the cosmic web in a similar manner.
\cite{massara_2021} shows that the marked power spectrum, conceptually similar to a density field transformation, sets greatly improved constraints on all cosmological parameters.

In the recent years, new methods applying non-linear operators on top of wavelet transforms, the so-called Wavelet Scattering Transform (WST) \cite{bruna2012invariant}, have also obtained promising results~\citep{chengwst2020,chengwst2021,valogiannis2021optimal}.
\cite{valogiannis2021optimal} for instance suggested that the WST can improve constraints on the value of the cosmological parameters by a factor between 3 and 100 better than the power spectrum, when evaluated on the 3D matter density field. 
A similar method called the Wavelet Phase Harmonics has been introduced in~\cite{Allys_2020}, showing very promising results in terms of information content\footnote{In this work, we use both information content and parameter constraints. Higher information content means tighter parameter constraints, and the other way around}.
It is a standard practice in cosmology to quantify the information content a given statistic carries by using the Fisher matrix formalism. For instance, the Quijote simulations \citep{Villaescusa_Navarro_2020}, a suite of 44100 full N-body simulations was designed to perform Fisher matrix calculations and several of the works listed above employ such simulations to address this point. 

Although conceptually simple, the standard Fisher matrix analyses rely on some assumptions like the Gaussianity of the considered statistic. In this work, we investigate the level of non-Gaussiantities of different statistics and their impact on Fisher matrix calculations. Overall, we argue how the use of several statistical tools can help in the quest to find optimal and robust statistics to extract the maximum information from the cosmic web and its tracers.

The rest of the paper is organized as follows:. 
\begin{enumerate}
    \item First, in Sec. \ref{sec:methods} we introduce the Fisher matrix formalism and two statistical tests to quantify the level of non-Gaussianity in a given statistic. We also propose a method to remove non-Gaussian components from the considered statistic.
    \item Second, in Sec. \ref{sec:Pk} we illustrate the problem by considering the power spectrum and some statistics derived from it and show how the Fisher matrix formalism can give different results just a result of transformations that do not carry cosmological information. We show how to ameliorate these situations by making use of the non-Gaussian tests.
    \item Third, we repeat the above exercise but for other statistics of the large-scale structure of the Universe such as the bispectrum, marked power spectrum, and WST in Sec. \ref{sec:nonGaussian}.
    \item Next, we describe the limitations of the tools used to identify non-Gaussianities in Sec. \ref{sec:hngc}.
    \item Finally, we draw our conclusions in Sec. \ref{sec:conclusions}.
\end{enumerate}

\section{The Fisher matrix formalism and Gaussianity tests}
\label{sec:methods}

In this section we first describe the Fisher matrix formalism and then we discuss two different tests to identify non-Gaussianities in a given statistics. We then describe a method to remove non-Gaussian dimensions from generic statistics. We note that while in this paper we focus our attention on cosmology, these methods are generic and can therefore be applied to problems outside cosmology.

\subsection{Fisher matrix Formalism}

The Fisher matrix formalism (\cite{FisherOnTM,Cover2006}) is a method to quantify the accuracy that a given statistic can constrain the value of some parameters. The Fisher matrix formalism is commonly used in cosmology to quantify the accuracy that a given statistic can place on the value of the cosmological parameters. One of its big advantages is that is does not require actual data to perform the calculation. 

When having N parameters, $\bm{\theta}\in \mathcal{R}^N$, conditioning the value of a statistic $\bm{X}$, the Fisher information can be represented in a matrix form as:
\begin{equation}
    F_{ij}(\bm{\theta})=E_{\bm{X}}\left[\left(\frac{\partial}{\partial \theta_i}\log~\mathcal{L}(\bm{X};\bm{\theta})\right)\left(\frac{\partial}{\partial \theta_j}\log~\mathcal{L}(\bm{X};\bm{\theta})\right)| \bm{\theta} \right],
\end{equation}
where $\mathcal{L}(\bm{X};\bm{\theta})$ is the likelihood of $\bm{X}$ conditioned on $\bm{\theta}$. When the likelihood can be differentiated twice, this can be rewritten as
\begin{equation}
    F_{ij}(\bm{\theta})=-E_{\bm{X}}\left[\frac{\partial^2}{\partial \theta_i \partial \theta_j}\log~\mathcal{L}(\bm{X};\bm{\theta})|\bm{\theta}\right].
\end{equation}
This matrix is called the Fisher information matrix (FIM) \cite{FisherOnTM,Cover2006} The Cramer-Rao theorem states that the variance of an optimal unbiased estimator on the parameter $\theta_i$ will satisfy
\begin{equation}
    \delta^2 \theta_i \geq (F^{-1})_{ii}~.
    \label{Eq:variance_FIM}
\end{equation}
When the likelihood $\mathcal{L}(\bm{X};\bm{\theta})$ is a multivariate Gaussian distribution, the Fisher information matrix can be expressed as \citep[see e.g.][]{tegmark1997}
\begin{equation}
    F^{\theta}_{ij}=\frac{\partial \mu_k}{\partial \theta_i}\frac{\partial \mu_l}{\partial \theta_j}\Sigma^{-1}_{kl}+\frac{1}{2} \Sigma^{-1}_{kl}\frac{\partial \Sigma_{lm}}{\partial \theta_i}\Sigma^{-1}_{mn}\frac{\partial \Sigma_{nl}}{\partial \theta_j},
\end{equation}
where $\mu$ and $\Sigma$ are the mean and the covariance of the considered statistic. In this equation and in the whole paper, we assume Einstein notation. Following~\cite{carron_2013}, we only keep the first term in this equation since the second one leads to overestimating the Fisher information in the assumption of a Gaussian likelihood. We do not come back on this hypothesis in the present paper. The Fisher matrix is then further simplified as:
\begin{equation}\label{eq:Fisher}
    F^{\theta}_{ij}=\frac{\partial \mu_k}{\partial \theta_i}\frac{\partial \mu_l}{\partial \theta_j}\Sigma^{-1}_{kl}~.
\end{equation}
To evaluate the FIM (e.g. from numerical simulations) two ingredients are needed:
\begin{enumerate}
    \item Estimate the covariance $\Sigma$ of the statistic, which can be computed from many independent realizations, at fixed value of the cosmological parameters, of the considered statistic.
    \item Estimate the partial derivatives of the expectation value of the statistic with respect to the parameters.
\end{enumerate}
In theory, this is enough to evaluate the FIM and to derive optimal constraints on the cosmological parameters from Eq. \ref{Eq:variance_FIM}. In practice, however, there are a few subtleties to this analysis, such as:
\begin{enumerate}
    \item The estimated covariance and/or derivatives might have not numerically converged.
    \item Numerical precision can affect calculation of derivatives and matrix inversion. 
    \item Spurious effects may arise due to artifacts from the way the statistic is represented.
    \item Noise and systematics may not have be taken into account.
    \item The likelihood of the considered statistic can be substantially non--Gaussian.
\end{enumerate}

It is common practice to perform some sanity checks to verify that the first and second points above are not a problem. There are also standard practices to investigate the effects of the third. While including noise may be easy, systematics maybe more challenging. 
In this work however, we focus our attention on the last point, that it is usually not taken into account and it is commonly assumed that the likelihood is a multivariate Gaussian distribution.

\subsubsection{Standard Fisher Analysis}

We will start with a \textit{standard} Fisher analysis, where we evaluate the Fisher matrix of Eq. \ref{eq:Fisher} and derive optimal constraints using Eq. \ref{Eq:variance_FIM}. In this analysis, we will perform a series of sanity check to verify the robustness and validity of the computation, such as:
\begin{itemize} 
    \item We check that the condition number\footnote{The condition number is defined as the ratio between the maximum and the minimum eigenvalue of a given matrix.} of the covariance matrix is well under $10^7$. Larger values can lead to numerical unstabilities when computing the inverse of the covariance matrix.
    \item We conservatively remove any frequency beyond $k_{Ny}$, the Nyquist frequency of the grid. 
    \item We check the numerical convergence of the covariance and the derivatives by checking the change in the constraints when using a subset of the simulations.
\end{itemize}

In some cases, we find that even if the estimate of the covariance of the statistic has converged at the percent level, its inverse might have errors at the unity level. To check this we repeat the whole analysis assuming we only had 70\% of the simulations for the covariance and the derivatives. We overplot these in every plot, labelled ``sub". 

\subsubsection{Fisher analysis from Quijote simulations}

In this paper, the different Fisher computations are carried out using the Quijote Suite, which is especially designed for this purpose. We consider six cosmological parameters, $\{ \Omega_m,\:\Omega_b,\:h,\:n_s,\:\sigma_8,\:M_{\nu}\}$ \citep[see][for the choice of cosmological models]{Villaescusa_Navarro_2020}. In particular, we use:

\begin{itemize}
\item A set of 15000 simulations with the same fiducial cosmology, closely matching the latest constraints by Planck \citep{planck2018}, to estimate the covariance matrix. 

\item A pair of 500 simulations ran with one parameter slightly smaller and bigger that is fiducial value to estimate the partial  derivatives of the statistic with respect to the parameters $\{ \Omega_m,\:\Omega_b,\:h,\:n_s,\:\sigma_8\}$. To compute the partial derivative of the statistic with respect to $M_{\nu}$, we instead use four sets of 500 simulations ran with $M_{\nu}=0.0,\:0.1,\:0.2,\:0.4$ eV neutrinos. 
\end{itemize}

The $M_\nu=0.0$ eV simulations have the same parameters as the fiducial simulations, but they have been generated from Zeldovich initial conditions as in the massive neutrino simulations. The value of the parameters for all the simulations employed can be found in Table \ref{tab:sims}. We refer the reader to \cite{Villaescusa_Navarro_2020} for further details on the Quijote simulations. 

\begin{figure}[t!]
\fig{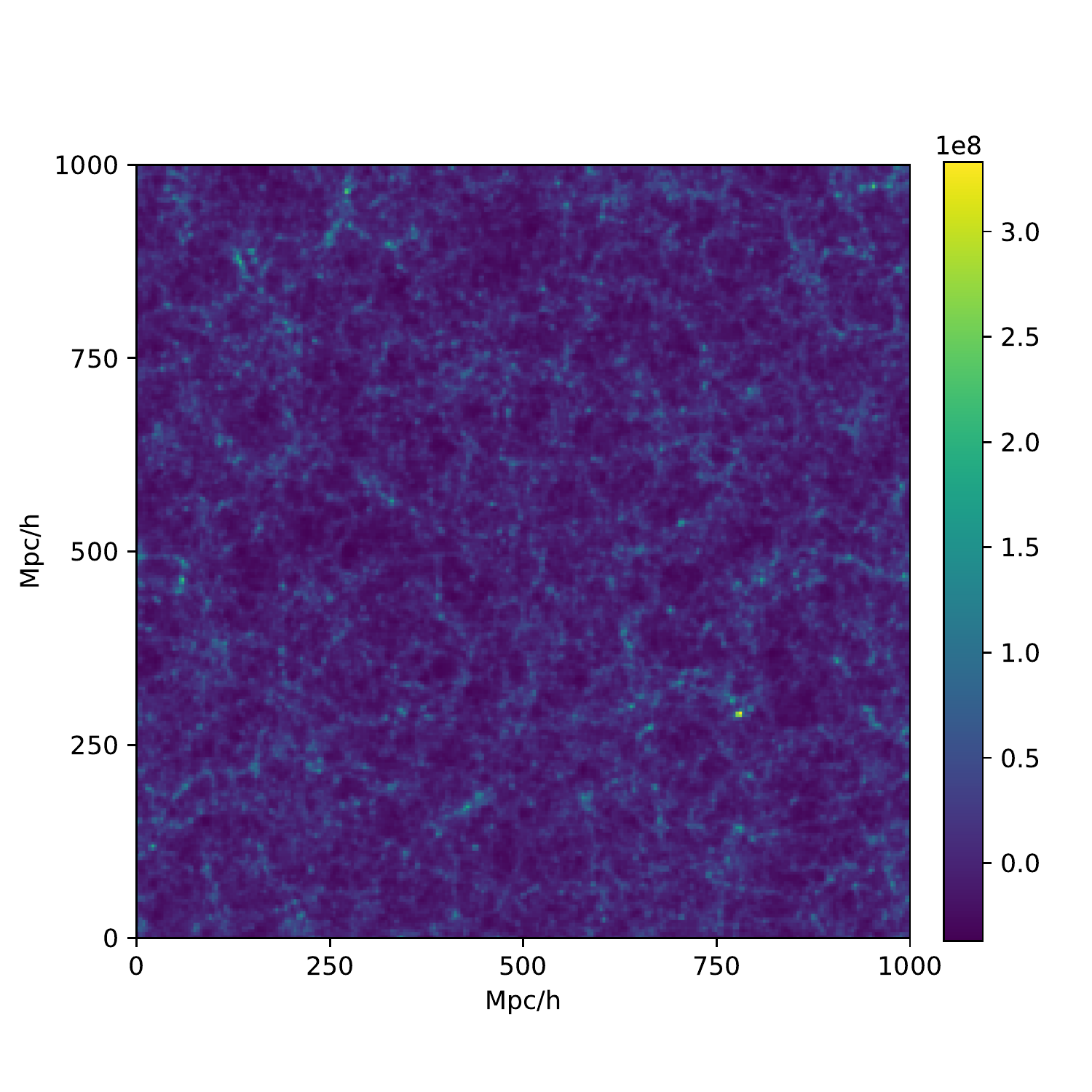}{.45\textwidth}{}
\caption{This figure shows an example of 2D density field from a Quijote simulation. The slice has dimensions of $1000\times1000\times250~(h^{-1}{\rm Mpc})^3$. As can be seen, these fields are non-Gaussian and therefore the power spectrum cannot characterize all of its statistical properties. \label{fig:LSSexample}}
\end{figure}

In this work we focus our attention on summary statistics of the 3D matter density field (see Fig. \ref{fig:LSSexample} as an example). In future work we plan to carry out this exercise for summary statistics of the halo and galaxy density fields. 

\subsection{Probing the high--dimensional non--Gaussianity of the statistic distributions}\label{sec:ngpair}
Probing the Gaussianity (normality) of a probabilistic variable can be done via many tests in 1D. For instance, a combination of the kurtosis and skewness yields a simple but efficient and fast descriptor for the non--Gaussianity~\citep{kurskew}, the Kolmogorov-Smirnov test \citep{stat_handbook} can evaluate the goodness-of-fit between empirical and expected cumulative distribution functions (CDFs), and the Shapiro--Wilk test \citep{Shapiro2007AnAO} is another efficient test to reject a null hypothesis about Gaussianity. 

However, the task becomes more complex and challenging in higher dimensions. In this work we will perform two tests, one to identify and quantify non-Gaussian pairs, and another to quantifying whether the sharpness of the likelihood is reproduced by the Gaussian assumption.
\subsubsection{Pairwise Gaussianity test}
\label{sec:pairwise}

For some applications, it may be interesting to quantify the Gaussianity of the different dimensions of an statistic. To identify the terms exhibiting non--Gaussianity, we use a simplified version of the test proposed in \cite{sellentin_2017}. The steps, nearly identical to \cite{sellentin_2017} are the following:
\begin{enumerate}
    \item Start with N samples of a d--dimensional statistic, $\mathbf{S}\in \mathcal{R}^{(N,d)}$, where the sample mean has been subtracted,  $\sum_b S_{bi}=0$.
    \item Compute the covariance:
    \begin{equation*}
        \mathbf{C}=\frac{\mathbf{S^T}\mathbf{S}}{N-d-2}
    \end{equation*}
    and check its convergence\footnote{Convergence here is checked by the percent level convergence of the covariance when using $80\%$ of the simulations. However, the convergence of the covariance does not guarantee the convergence of its inverse or any derived quantities. As we will see, we use mock data to overcome these difficulties.}. Note that the denominator includes the Hartlap factor \citep{hartlap_2006}\footnote{If we were to omit this factor, the mean of $t_b$ (defined in the next item) would be away from the expected mean, $d$.}. 
    
    \item For all $(i,j)$ such that $0\leq i,j <d$ and $\: i\neq j$, get the two eigenvectors $\vec{v}_{(i,j)},\:\vec{w}_{(i,j)}$ of the sub --covariance matrix
    \begin{equation}
    \begin{pmatrix}
        C_{ii} & C_{ij} \\
        C_{ji} & C_{jj}
    \end{pmatrix}
    \end{equation}

    \item For all $0\leq b <N$ and all pairs $(i,j)$ calculate:
    \begin{align*}
        x_{bij}&=\vec{v}_{(i,j)}\cdot (S_{bi},S_{bj})\\
        y_{bij}&=\vec{w}_{(i,j)}\cdot (S_{bi},S_{bj})
    \end{align*}
    Now, if $\mathbf{S}$ are samples from a multivariate Gaussian, for each $(i,j)$,
    \begin{align*}
        z_{bij}=x_{bij}+y_{bij}
    \end{align*}
    should be samples drawn from a Gaussian as well.

    \item Perform a kurtosis--skewness test \citep{kurskew} on $z_{bij}$ for all $(i,j)$ along $b$ and construct the matrix:
    \begin{align}\label{pairwise_sum}
        R_{ij}=s_{ij}^2+k_{ij}^2
    \end{align}
    where s is the z--score from the skewness test and k is the z--score from the kurtosis test, both along the sampling dimension. See Eqns. 13 and 19 of \cite{skew} for reference.

\end{enumerate}
We will refer to this test as the pairwise Gaussianity test.
The p--value for the test in Step 5 can also be of interest, but this is prone to numerical error and stochastic convergence, so we rather choose to run many calibrations using the covariance obtained in Step 2. 
We draw samples from a  multivariate Gaussian having the covariance estimated in Step 2 and repeat the Gaussianity test with these samples. 
We perform the same tests above with this mock data. We denote the mean of $R_{ij}$ over different mocks as $\mu^{cal}_{ij}$ and the standard deviation of $R_{ij}$ over different mocks as $\sigma^{cal}_{ij}$. ``cal" stands for ``calibration".

\subsubsection{Quantifying the overall non--Gaussianity}
\label{sec:ngchisq}

The second test we use to quantify the level of non-Gaussianity of an statistic evaluates how well a multivariate Gaussian approximate the shape of the likelihood around the fiducial parameters. In general, this test works well when there are enough samples to obtain a converged estimate of the covariance matrix. Our test for a s--sigma confidence level is described below. The index $b$ always runs over the different samples while $i,\:j$ runs over the dimensions of the statistics:
\begin{enumerate}
    \item Start with N samples of a d--dimensional statistic, $\mathbf{S}\in \mathcal{R}^{(N,d)}$, where the sample mean has been subtracted,  $\sum_b S_{bi}=0$.
    \item Divide $\mathbf{S}$ into two sets of $N/2$ samples. We denote the first set as $\mathbf{A}\in \mathcal{R}^{(N/2,d)}$ and the second set as $\mathbf{B}\in \mathcal{R}^{(N/2,d)}$
    \item Compute the covariance using only $\mathbf{A}$:
    \begin{equation*}
        \mathbf{C}=\frac{\mathbf{A^T}\mathbf{A}}{N/2-d-2}
    \end{equation*}
    and check the convergence of the matrix elements by using smaller ($<N/2$) number of samples.
    
    \item Evaluate $t_b=B_{bi}\:C_{ij}\:B_{bj}$ (no sum on b). The square root of this quantity is also called the Mahalanobis distance.
    
    \item If the statistic distribution is Gaussian, the $t_b$ values are expected to follow the $\chi^2$-distribution for $d$ degrees of freedom. 
    
    \item Use the Komogorov--Smirnov test \citep{stat_handbook} of these $t_b$ values and the $\chi^2$-distribution of degree of freedom $d$. We get the test statistic:
    \begin{equation*}
        s_{\rm KS}={\rm sup}_x|{\rm CDF}_{t_b}(x)-{\rm CDF}_{\chi^2_d}(x)|~,
    \end{equation*}
    where ${\rm CDF}_{t_b}$ is the empirical CDF from the $t_b$ samples and ${\rm CDF}_{\chi^2_d}$ is the CDF of the $\chi^2$-distribution of degree of freedom $d$. Note that these CDFs are 1--dimensional.

    \item Repeat with some mock samples drawn from a Gaussian with the covariance obtained in Step 3. The test passes if the test statistic, $s_{\rm KS}$, is within a s--sigma interval from the Gaussian mock. In this work we use $s=3$ and $s=5$.
\end{enumerate}

We note that different metrics can be used to evaluate the distribution differences in Step 6. We tested out some options including the Kullback--Leibler divergence and the Earth mover's distance, but found them to be more sensitive to the outlier samples at the tail of the distribution. We call this test the $\chi^2$ distributional test.

With the two Gaussianity tests described above, we aim at identifying two signatures of a non--Gaussian likelihoods: 1) when pairs of coefficients shows a highly non--Gaussian relation, and 2) when the overall likelihood peak's sharpness differs from the Gaussian one. We will use these two tests to quantify, and remove, the non-Gaussianities of different statistics of the large scale structure.

\subsection{Removing the non--Gaussian dimensions}
\label{sec:gaussianizing}

Based on the above analysis, we propose a scheme to iteratively eliminate the  non-Gaussian components of a given statistic, keeping a subset that passes our Gaussianities tests at some confidence level. The procedure is as follows

\begin{enumerate}
    \item Compute $R_{ij}$, $\mu^{cal}_{ij}$ and $\sigma^{cal}_{ij}$ for all $(i,j)$ based on Equation \ref{pairwise_sum}.
    
    \item Perform the pairwise Gaussianity test
    \begin{enumerate}
        \item Compute the matrix of z--scores $Z_{ij}$ where
        \begin{align}\label{eq:z}
            Z_{ij}&=(R_{ij}-\mu^{cal}_{ij})/\sigma^{cal}_{ij}
        \end{align}
    
        \item In order to remove the maximally non Gaussian component, remove the row containing the maximal matrix element of $Z_{ij}$. Since we would get two rows, we remove the row in which the sum of the $Z_{ij}$ along the row is bigger.
    
        \item Repeat (b) until all z--scores lay within a s--sigma interval.
    \end{enumerate}
    
    \item Perform the $\chi^2$ distributional test
    \begin{enumerate}
        \item Compute $z_i=\sum_j Z_{ij}$.
        
        \item Eliminate dimensions sorted by decreasing value of $z_i$ until the remaining statistic passes the $\chi^2$ distributional test within a s--sigma interval
    \end{enumerate}
    
    \item The remaining statistic are the dimensions surviving both tests.
\end{enumerate}
We will refer to ``\textit{Gaussianize} a given statistic'' when we apply to it the above procedure. It is however important to emphasize that this does not mean that we take a non-Gaussian statistic and make it Gaussian, but instead that we attempt to remove its non-Gaussian components. Thus, this procedure will naturally remove information from the statistic.

\begin{figure*}[t!]
\gridline{\fig{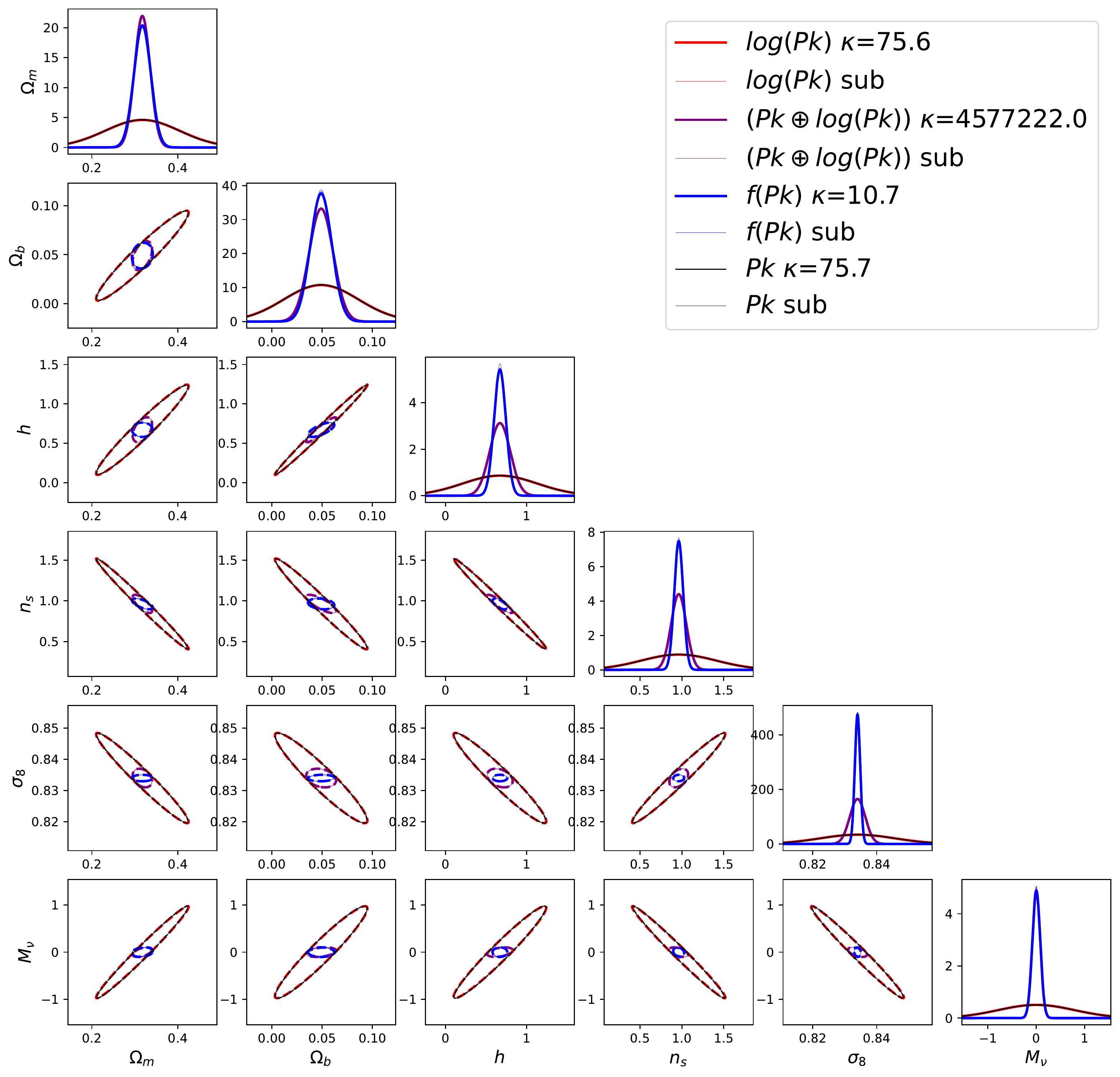}{.45\textwidth}{(a) Fisher constraints for $Pk,\:\log(Pk),\:Pk\oplus \log(Pk),\:f(Pk)$.}
\fig{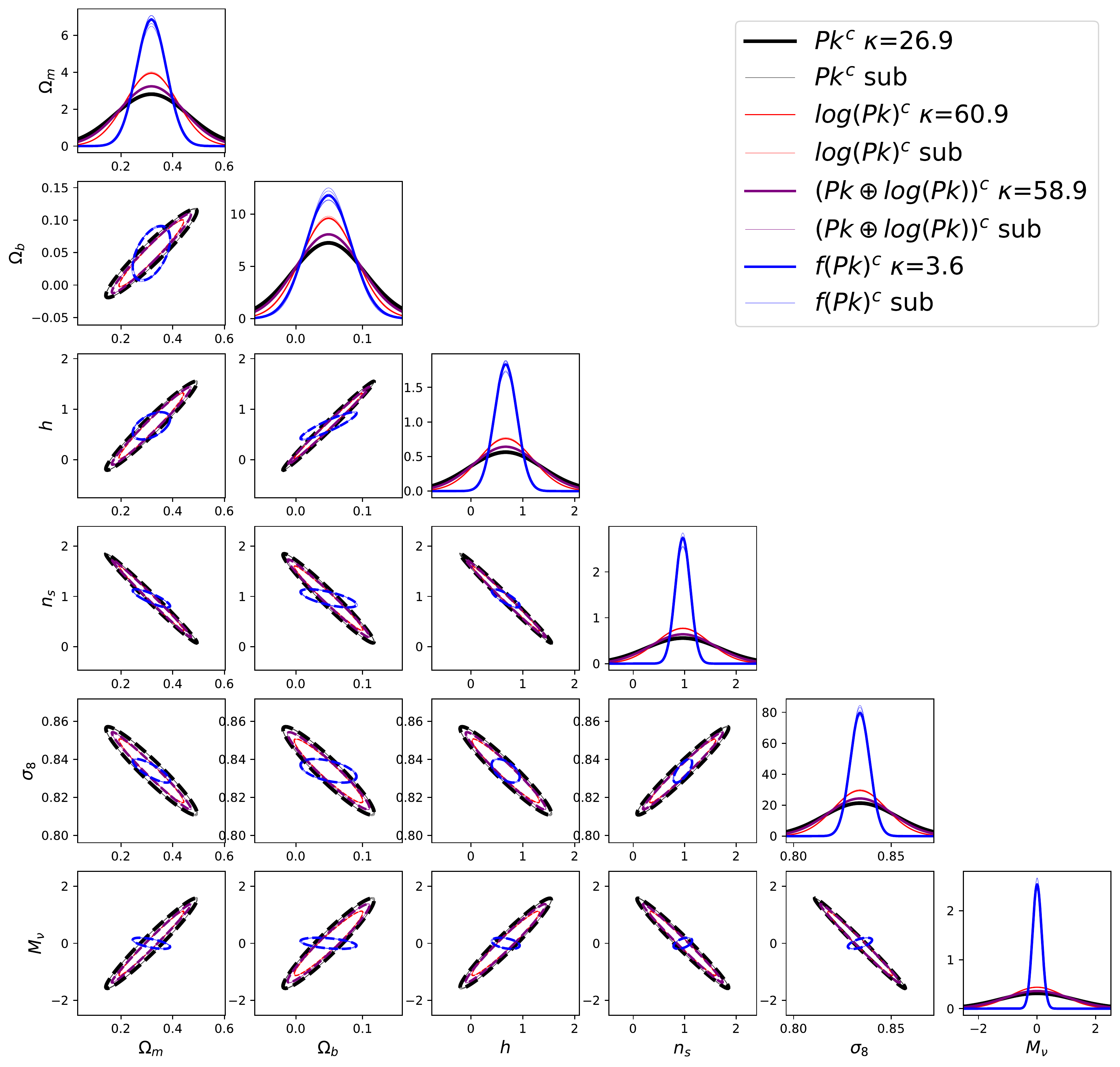}{.45\textwidth}{(b) Fisher constraints for \textit{Gaussianized} $Pk,\:\log(Pk),\:Pk\oplus \log(Pk),\:f(Pk)$.}
}
\caption{We have used the Fisher matrix formalism to quantify how well a given statistic can constrain the value of the cosmological parameters. The left panel show the results for $Pk,\:\log(Pk),\:Pk\oplus \log(Pk),\:f(Pk)$, while the right panel show the same for the Gaussianized equivalent (i.e. the statistic obtained after removing the non-Gaussian components as explained in Sec. \ref{sec:gaussianizing}). ``sub'' refers to results obtained 70\% of the data, and $\kappa$ denotes the value of the conditional number. As can be seen, $Pk\oplus\log Pk$ and $f(Pk)$ achieves tighter constraints on the value of the parameters than $Pk$ and $\log Pk$ (they achieve similar constraints), which is not possible. Their \textit{Gaussianized} version achieve constraints much more similar. We note however that for $f(Pk)$ we were not able to keep enough Gaussian dimensions to obtain reliable Fisher constraints.
This exercise shows the importance of quantifying and avoiding using non-Gaussian statistics using traditional Fisher matrix calculations.
\label{fig:pklogpkfpk}}
\end{figure*}

\section{Examples with the power spectrum and its variations}
\label{sec:Pk}

We now quantify how the constraints on the value of the cosmological parameters, as derived by a Fisher matrix computation, depend on the non-Gaussianity of the considered statistic. For this, we use the power spectrum and two toy statistics that are constructed from it.

\subsection{Statistical probes}

We start by describing the power spectrum and the two toy statistics we build from it.

\subsubsection{The Power Spectrum ($\rm{Pk}$)}\label{sec:pk}

The power spectrum characterizes the amplitude of Fourier modes for different wavenumbers. For an homogeneous and isotropic random field, $\delta(\mathbf{x})$, one can define the (isotropic) power spectrum as
\begin{equation}\label{eq:pk}
    \langle \tilde{\delta}(\mathbf{k}) \tilde{\delta}^{*}(\mathbf{k'}) \rangle=(2\pi)^3P(k)\delta_D^3(\mathbf{k}-\mathbf{k'})
\end{equation}
where the brackets indicate an ensemble average, $\delta(\bf{k})$ is the Fourier transform of $\delta(\bf{x})$, and $\delta_D^3$ is a Dirac delta.
Being an isotropic estimator, it depends only on the norm $k$ of $\mathbf{k}$, the only non-vanishing configurations being for $\mathbf{k}=\mathbf{k'}$.
The power spectrum, as a probe of the LSS, has the advantage of being directly interpretable and closely related to theoretical predictions.
For an isotropic and homogeneous Gaussian random field, the power spectrum contains all the information about the underlying process. 
Indeed, all the odd higher--order correlation functions vanish and the even correlation functions can be expressed as functions of the power spectrum. 

It is worth mentioning that the power spectrum of a non--linear transformation of the density field has been shown to be a useful statistic for cosmology. For instance, the power spectrum of the log of the density field \citep{neyrinck_2009} and  the clipped power spectrum \citep{simpson_2011,simpson_2013} are examples of statistics that bring information from high-order correlation functions back to the power spectrum due to the non-linear field level transformation.

We have performed a standard Fisher matrix analysis using the power spectrum and show the results in Fig. \ref{fig:pklogpkfpk} with black lines. We also show results for the convergence tests that we denote as ``sub''. We find that results passes all standard tests: small conditional number and convergence for covariance and derivatives.

\subsubsection{$\rm{Pk} \oplus \rm{log(Pk)}$}
\label{sec:pksqrtpk}

We will illustrate the problem of performing Fisher matrix analysis using non-Gaussian statistics by constructing a toy statistic whose likelihood is not Gaussian. We consider the statistics defined by the concatenation of the power spectrum, $\rm{Pk}$, and the log of the power spectrum, $\rm{log(Pk)}$. We denote this statistics as $\rm{Pk} \oplus \rm{log(Pk)}$. 

In Fig. \ref{fig:pklogpkfpk} we show the derived constraints on the value of the cosmological parameters from a standard FIM for $\rm{Pk}$, $\rm{log(Pk)}$, and $\rm{Pk}\oplus\rm{log(Pk)}$. As can be seen, the constraints from the $\rm{Pk}\oplus\rm{log(Pk)}$ are tighter than the ones from $\rm{Pk}$ and $\rm{log(Pk)}$ (being these two very similar). This is physically not possible, since we are just performing a local transformation of the power spectrum, that cannot add additional information to the existing one from the power spectrum.

One could think that this behaviour may be happening because $Pk$ and $\log Pk$ are very correlated, and that computing properly their cross-covariance will get the correct results. However, this is not what we found since our standard Fisher analysis passes all traditional tests to determine the robustness of the results.

\begin{figure*}
\gridline{\fig{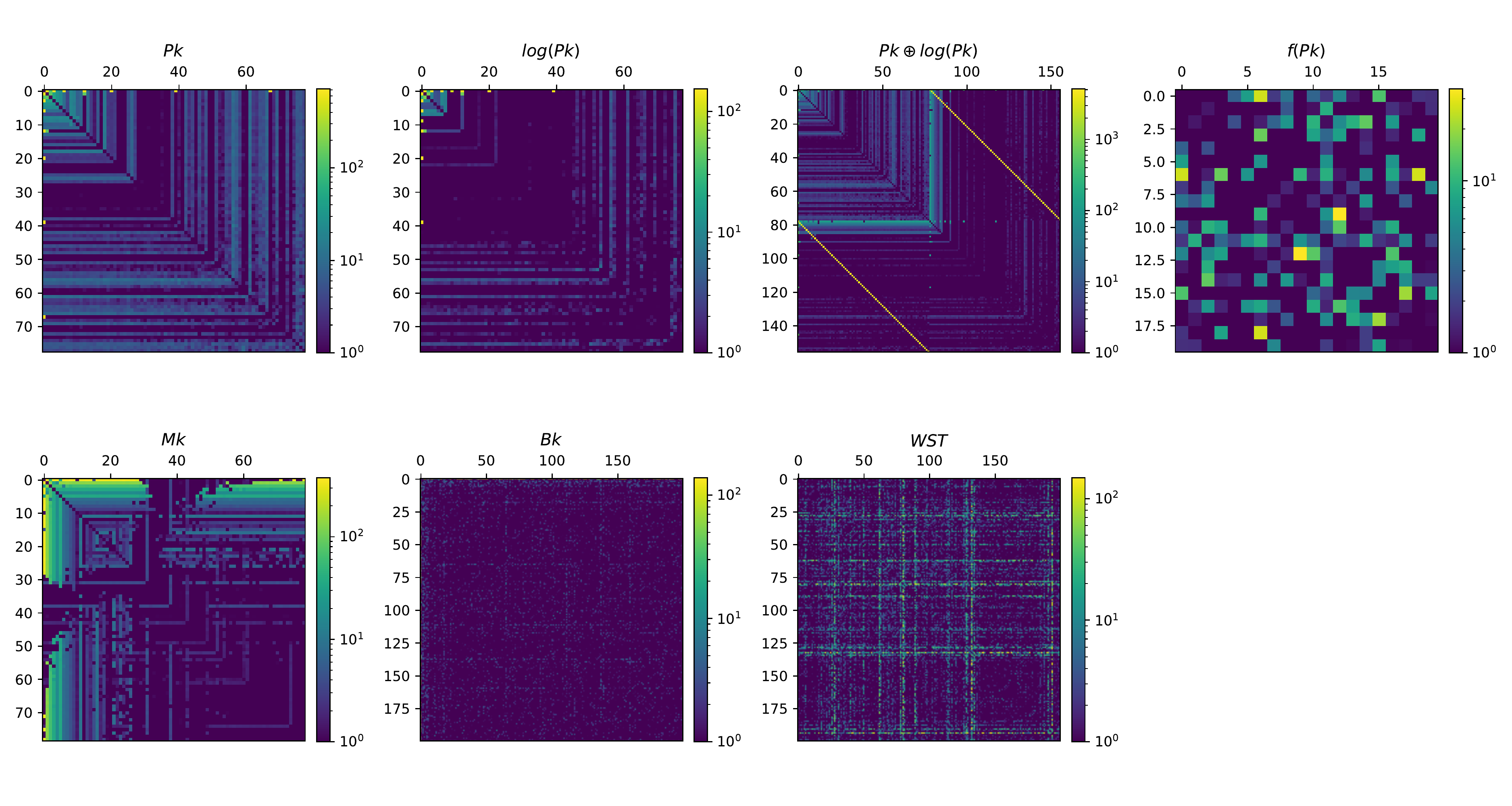}{1\textwidth}{}}
\caption{We have performed the pairwise non--Gaussianity test on a set of different statistics: $\rm{Pk}$, $\rm{log(Pk)}$, $\rm{Pk}\oplus\rm{log(Pk)}$, $f(\rm{Pk})$, $\rm{Mk}$, $\rm{Bk}$, ${\rm WST}$ (from top-left to bottom right). The color in each pixel indicates the z--scores, $Z_{ij}$, defined in Equation \ref{eq:z}. Higher values indicate larger deviations from Gaussianity. We find different patterns in the pairwise non--Gaussianity matrices. Note that $\rm{Pk}$, $\rm{log(Pk)}$, $\rm{Mk}$ are ordered such that the large scales (small $k$) comes first. The bright bands around the 80\textsuperscript{th} element of $\rm{Pk}\oplus\rm{log(Pk)}$, are pairs between the large scales of $\rm{log(Pk)}$ and all scales of $\rm{Pk}$. The bispectrum, $\rm{Bk}$, and wavelet scattering transform, ${\rm WST}$ are reduced to 200 dimensions for the ease of analysis. This test can help us identifying and removing non-Gaussian components of a given statistic. \label{fig:ngmat}}
\end{figure*}

\subsubsection{Arbitrary Transformation of the Pk: $f(\rm{Pk})$}

We now show another example of an statistic derived from the power spectrum that can give rise to unrealistically tight constraints on the value of the cosmological parameters.

We build the summary statistic, that we call $f(\rm{Pk})$, as follows. We optimized a multi layer perceptron (MLP) network that takes as input the power spectrum and outputs a non-linear function of it. We minimize a loss function that represents the parameter constraints derived from a standard FIM. Specifically, we use a MLP with 2 hidden layers with 32 neurons which transforms the 78 dimensional power spectrum into a 10 dimensional statistic. 
We apply the ReLU activation function \citep{relu_citation} to the output of each hidden layer.
Let the parameters of the network be $\bm{\lambda}$, then we optimize for
\begin{equation}
    \bm{\lambda'}=\rm{argmax}_{\bm{\lambda}}\:\mathcal{L}(\bm{\lambda})~,
\end{equation}
where 
\begin{equation}
    \mathcal{L}(\bm{\lambda})=\mathcal{L}_{\Delta\theta}(\bm{\lambda})+\mathcal{L}_{\rm NG}(\bm{\lambda})+\mathcal{L}_{\rm Cond}(\bm{\lambda})~.
\end{equation}
$\mathcal{L}_{\Delta\theta}(\bm{\lambda})$ is the loss term decreasing the marginalized parameter constraints. It is implemented as the sum of the squares of the ratio of the new constraint to the constraint given by $\rm{Pk}$. 
$\mathcal{L}_{\rm NG}(\bm{\lambda})$ is the loss term maintaining the statistic to be dimension--wise Gaussian. 
It is simply $({\rm Skewness})^2+({\rm kurtosis}-3)^2$.
$\mathcal{L}_{\rm Cond}(\bm{\lambda})$ is just the condition number of the covariance when using this statistic. 
See Appendix \ref{app:fpk} for further details on the loss function and its different terms.
Then our statistic becomes $f(Pk)={\rm MLP}(Pk,\bm{\lambda'})$. We show the parameter constraints, derived from the FIM in Fig. \ref{fig:pklogpkfpk}. As in the case of $Pk\oplus \log Pk$, $f(\rm{Pk})$ achieves higher accuracy on the cosmological parameter than the power spectrum. This is physically non possible as both statistics are related by a transformation that does not contain cosmological information.

\subsection{Non--Gaussianity tests}

To investigate whether the results above are due to their likelihood not being Gaussian we perform a pairwise Gaussianity test on $\rm{Pk}$, $\log Pk$, $Pk\oplus \log Pk$, and $f(\rm{Pk})$ and show the results in the upper row of Fig. \ref{fig:ngmat}. 

For the power spectrum, we find non-negligible non--Gaussianities at the largest scales. This is expected since on large scales there are few modes and the power spectrum is not expected to follow a Gaussian distribution. This observation is somewhat similar to the one in \cite{sellentin_2017} for the weak lensing power spectrum. 
We also find some non-Gaussianities on small scales. However, we suspect this is due to numerical artifacts when calculating the power spectrum.

For the logarithm of the power spectrum, we find significantly lower non--Gaussianities, although we observe some on large scales. In this case, since the power spectrum spans several orders of magnitude, we believe that a logarithmic transform could in part reduce the effect of outliers on the covariance. For $\rm{Pk}\oplus\rm{log(Pk)}$, we observe that the non--Gaussianity between a dimension of $\rm{Pk}$ and the corresponding dimension of $\rm{log(Pk)}$ is clearly revealed by the pairwise Gaussianity test. For $f(\rm{Pk})$, we observe some pairs with non-negligible values of the z-score.

\begin{figure*}
\gridline{\fig{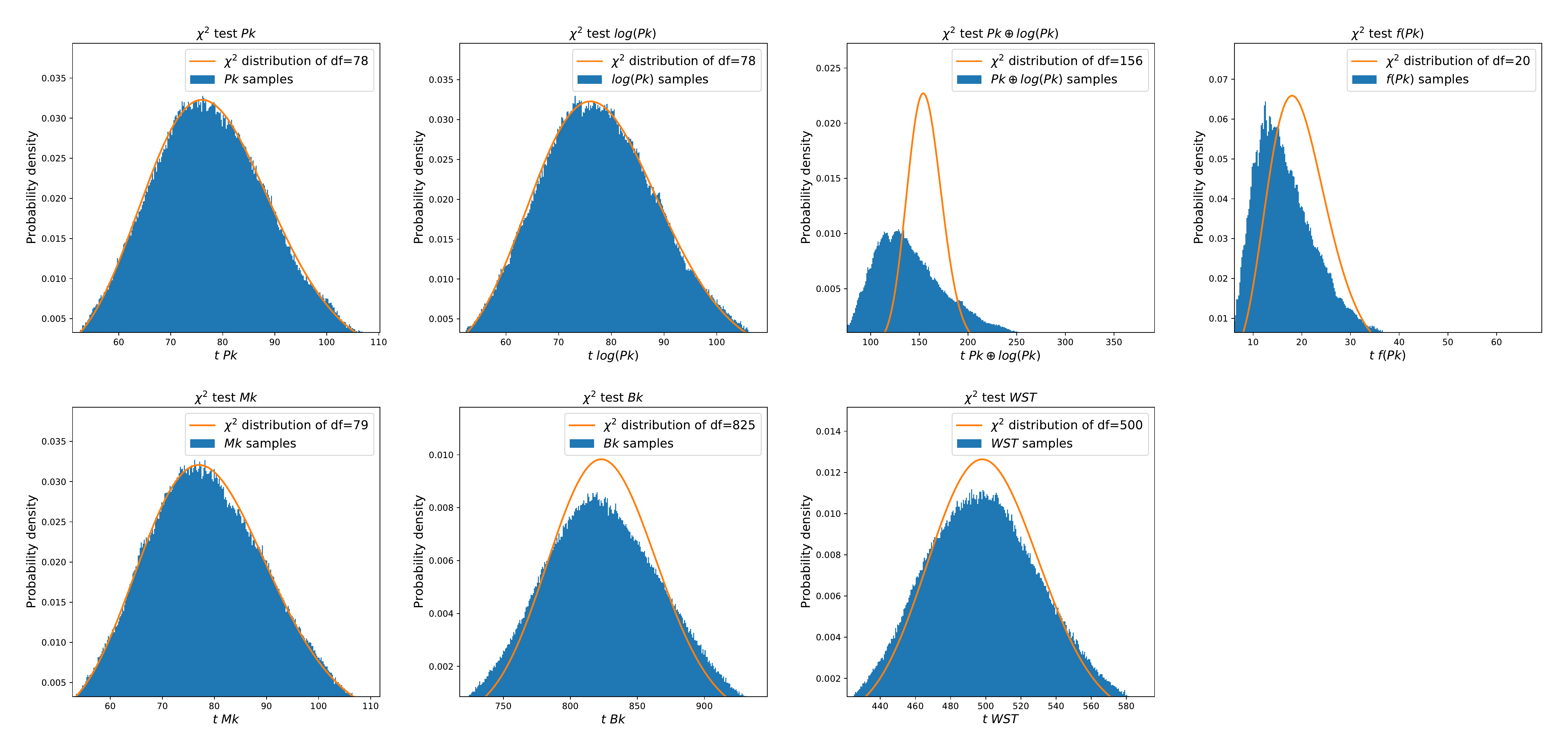}{.9\textwidth}{}}
\caption{Results of the $\chi^2$ distributional test performed on the seven statistics considered in this work:  $Pk,\:\log(Pk),\:Pk\oplus \log(Pk),\:f(Pk),\:Mk,\:Bk,\:{\rm WST}$. As can be seen, this can help us in identifying the statistics that deviate from Gaussianiy. In this case, $Pk\oplus \log(Pk),\:f(Pk),\:Bk,\:{\rm WST}$ exhibit different levels of non--Gaussian likelihoods.\label{fig:chisqs}}
\end{figure*}

We also perform the $\chi^2$ distributional test and show the results in the first row of Figure \ref{fig:chisqs}. While the CDF of $t$ values of $\rm{Pk}$ and $\rm{log(Pk)}$ shows negligible amount of deviation from the expected $\chi^2$ distribution, for $\rm{Pk}\oplus\rm{log(Pk)}$ and $f(\rm{Pk})$ we find substantial deviation from the expected distribution. $f(\rm{Pk})$, which was constrained to be dimension--wise Gaussian turned out to be highly non--Gaussian and does not pass the $\chi^2$ test. It is interesting to see that the pairwise non--Gaussianity test did not reveal these non Gaussianities as well as it did for other probes. The reason probably lies in the way we constructed the statistic. The output of a neural network is derived from dense linear operations and non--linearities. Thus its output coefficients can be expected to have correlations involving many terms compared to other probes which usually maintain some separation between the regions of Fourier plane which are probed. 
Even if we do not see much pairwise non--Gaussianities, it is likely that higher ($>2$) dimensions are correlated in a complex and non Gaussian manner.

The above tests indicate that the results from the standard Fisher matrix calculation for the $\rm{Pk\oplus\log Pk}$ and $f(\rm{Pk})$ may not be valid since these statistics exhibit significant level of non-Gaussianities.

\subsection{Corrected Fisher Analysis}

We now \textit{Gaussianize} the statistics using the procedure described in Sec. \ref{sec:gaussianizing} and show the results of the FIM analysis on the right panel of Figure \ref{fig:pklogpkfpk}. We refer the reader to Tables \ref{tab:change_table} and \ref{tab:change_r_table} for more quantitative details.

Although the non--Gaussianity detected for the power spectrum seems to be mild compared to the other probes, it does affect the parameter constraints at roughly a 50\% level, as we can see from Table \ref{tab:change_r_table}. We note however that this may be due to the fact that some of the wavenumbers identified as non-Gaussian on small scales may only be due to numerical artifacts. 

For $\rm{log(Pk)}$, we find that a logarithmic transform of the power spectrum is sufficient to make it more consistently Gaussian. The corrected parameter constraints are now only corrected at a 15\% level. It is important to emphasize that even if $\rm{log(Pk)}$ is just a transformation of the power spectrum, and therefore it should not contain more information that the power spectrum itself, the reason why our results show that constraints from the Gaussianized $\rm{log(Pk)}$ are better than those from the Gaussianized $\rm{Pk}$ is because our procedure removes non-Gaussian information. If that would not be the case, all statistics should give the same constraints.

For $\rm{Pk} \oplus \rm{log(Pk)}$, we observe that the non--Gaussianity between a dimension of $\rm{Pk}$ and the corresponding dimension of $\rm{log(Pk)}$ is clearly revealed. After correcting for the non--Gaussianity, $\rm{Pk} \oplus \rm{log(Pk)}$ ends up having constraints similar to that of $\rm{log(Pk)}$. We conclude that the non--Gaussian correlations and the spurious constraints caused by them are successfully removed.

\begin{figure*}
\gridline{\fig{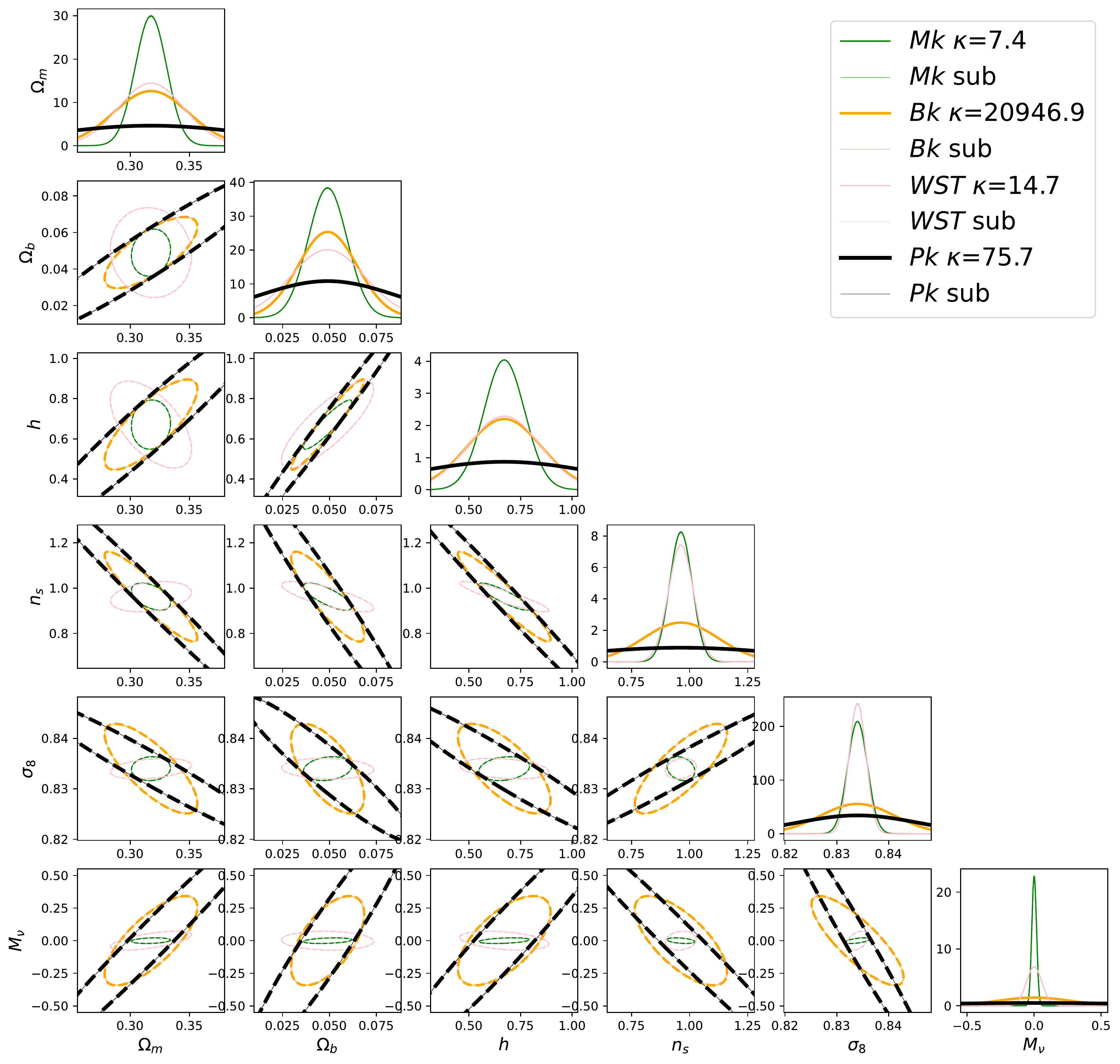}{.45\textwidth}{(a) Standard constraints for $Pk,\:Mk,\:Bk,\:WST$}
\fig{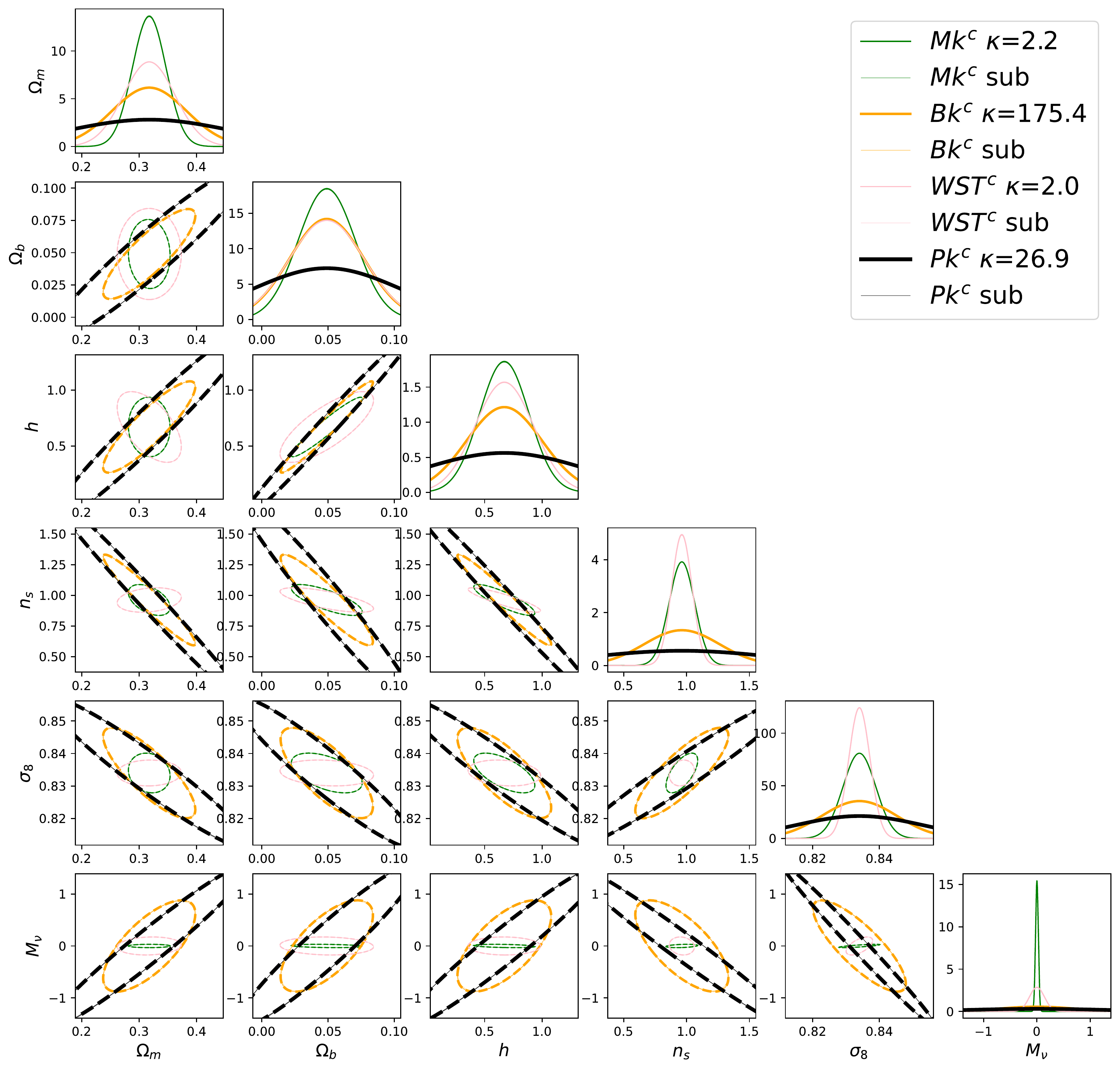}{.45\textwidth}{(b) Corrected constraints for $Pk,\:Mk,\:Bk,\:WST$}
}
\caption{Same as Fig. \ref{fig:pklogpkfpk} but for the marked power spectrum ($Mk$), bispectrum ($Bk$), and wavelet scattering transform (${\rm WST}$).
\label{fig:pkmkbkWST}}
\end{figure*}

For $f(\rm{Pk})$, and unlike other statistics, we find difficult to get consistent results when repeating the neural network training, or when bootstrapping the mock samples in the Gaussianity tests. It is also the case that the 70\% sub--runs tended to deviate more than the other statistics. In general, it should be thought to be unreliable. Although our test reveals that this statistic is exploiting the Gaussian assumption of the Fisher analysis to report seemingly confident results, we can see that the resulting ellipses are still quite promising from Figure \ref{fig:pkmkbkWST}, especially for the case of the neutrino mass. We do suspect that these constraints are still contaminated by other assumptions made for a Fisher analysis, and does not signifies that a function of the power spectrum can truly be more informative. We will try to reveal the cause in a future study. This observation however suggests that a spurious probe reporting seemingly confident results could be easily engineered while being hard to check if it is valid.

\section{Application to non-Gaussian statistics in cosmology}
\label{sec:nonGaussian}

In the previous section we have illustrated the problems inherent to estimating parameter constraints using Fisher matrix calculation for statistics that exhibit some level of non-Gaussianities. In this section we investigate the level of non-Gaussianities in statistics commonly employed to extract information not captured by the power spectrum such as the marked power spectrum, the bispectrum, and wavelet scattering transform (WST). We will also study the change in the Fisher results when we Gaussianize those statistics.

\subsection{Non--Gaussian statistics}

We now describe the different summary statistics we consider in this section. It is important to emphasize that the name of these statistics (non-Gaussian) does not arise due to their non-Gaussian distribution, but instead to the fact that they are used to study non-Gaussian density fields, where the power spectrum is not able to fully characterize its statistical properties. The likelihood of these statistics can still be Gaussian. 

The constraints on the value of the cosmological parameters derived from a standard Fisher analysis are shown in the left panel of Fig. \ref{fig:pkmkbkWST}.

\subsubsection{The marked power spectrum ($\rm{Mk}$)}

The idea behind the marked power spectrum is to assign a weight to each particle (or galaxy). That weight, can be an intrinsic property of the particle/galaxy or can be related to the environment of the object. 

In the cosmological context, the mark introduced by \cite{white_2016} has been studied in depth in \cite{massara_2021,philcox_2020}, especially for its ability to constrain the neutrino mass. %
In this work we use the measurements from \cite{massara_2021}. The mark here, firstly introduced in \citep{white_2016}, represents an environmental property of the particle/galaxy defined as
\begin{equation}\label{eq:mark}
    m(\vec{x};R,p,\delta_s)=\left[\frac{1+\delta_s}{1+\delta_s+\delta_R(\vec{x})}\right]^p~,
\end{equation}
with parameters $R = 10\:h^{-1}\rm{Mpc}$, $p = 2$, and $\delta_s = 0.25$.

\subsubsection{The Bispectrum ($\rm{Bk}$)}

The bispectrum is a statistics that measures correlations of closed triangles in Fourier space. For an homogeneous random field, it is defined as:
\begin{align}\label{eq:bk}
    \langle \tilde{\delta}(\mathbf{k_1}) \tilde{\delta}(\mathbf{k_2}) \tilde{\delta}(\mathbf{k_3}) \rangle &=\\ \nonumber
    (2\pi)^3&B((\mathbf{k_1},\mathbf{k_2},\mathbf{k_3})\delta_D^3(\mathbf{k_1}+\mathbf{k_2}+\mathbf{k_3}),
\end{align}
with the same notation of Eq.~\eqref{eq:pk}. Note that the bispectrum, as defined above, is a scalar function with three vector arguments. 
However, the delta function requires $\mathbf{k_1}+\mathbf{k_2}+\mathbf{k_3}=0$, i.e. the three vectors should form a triangle. Thus the bispectrum can also be represented as $B(k_1,k_2,\theta_{12})$ or $B(k_1,k_2,k_3)$ assuming statistical isotropy of the field. 
The bispectrum is a non--Gaussian statistic capturing interactions of different Fourier modes. In fact, the expectation value for the bispectrum vanishes for an homogenous Gaussian random field.
Recently, \cite{Hahn_2020} showed that the halo bispectrum is a good probe of Large Scale Structure breaking the parameter degeneracy between $\sigma_8$ and the sum of the neutrino mass $M_{\nu}$.

We use our own estimator for the bispectrum, which relies of Fast Fourier Transforms (FFT), similarly to other works \citep{sefusatti_2005,watkinson_2017}. We provide further details on the in the Appendix \ref{app:bk}.

\subsubsection{The Wavelet Scattering Transform (WST)}
The Wavelet Scattering Transform (WST) is a set of statistics initially used in image analysis. They were firstly introduced in \cite{bruna2012invariant,mallat2012group}. 
There are many similarities between WST and convolutional neural networks (CNNs) \citep{alexnet}, since they are both built from successive applications of convolutions and non-linearities. 
However, in the WST formalism, the convolutional kernels are a set of fixed wavelets instead of being optimized for the data, while the non-linearities are complex modulus. 
Wavelets are spatially localized oscillatory functions, which probe specific frequencies and orientations. 
Having a set of $N_f$ such wavelets which sample the whole Fourier space below the Nyquist frequency,
the wavelet transform of a field $I(\vec{x})$ is built by convolving it with these wavelets. This generates $N_f$ fields, which are bandpass filtered version of the original field on the frequencies probed by each wavelet.
The WST is then built with successive application of these wavelet convolutions and non-linear modulus operations, allowing to characterize the interaction between different frequency components of the field~\citep{mallat2012group}. Following recent works on the WST, we restrict ourselves to a two-layer WST. 
Recently, the WST became a statistic of interest in astrophysical applications \citep{allys_RWST_ISM_2019,rwst_Blancard_2020,saydjari_2021,cheng_new_2020,cheng_wstnumass_2021,cheng2021quantify}. 
In the present paper, to allow a direct comparison to other 3D statistics, we develop a ``2.5D" WST, where instead of using fully three dimensional wavelets, we treat the line of sight(LOS) direction specially. 
We dissect the xy--Fourier plane using radial and angular wavelets as in conventional 2D WST, but then we multiply each of the xy--wavelets by every other z--wavelet. 
Our z-wavelets are simply logarithmically spaced 1D wavelets in the z--direction. 
Our wavelets are thus not optimized to probe spherically isotropic fields but rather for a field with the LOS direction being special. 
This design of these wavelets might not be optimal as a statistic for an isotropic density field but is motivated by the fact that the line of sight direction is treated differently in real surveys. 
In this study, we use 2D wavelets with 8 angular bins and 8 radial bins and LOS(z) wavelets with 6 bins. We thus have $N_f=(1+8\times8)\times6=390$ wavelets and standardly $2+N_f+N_f^2=152492$ coefficients. However, we can average over angles since we assume statistical isotropy, and we assume that a convolution of a low passed image by a high frequency filter has negligible information \cite{mallat2012group}. We thus result with a probe with 1052 dimensions. Since our Gaussianity tests are computationally intensive for high dimensional probes, we further reduce the dimensionality to 500 dimensions by using a Principal Component Analysis. More details are in Appendix \ref{app:WST}.

\subsection{Results of the non--Gaussianity tests}

We have performed the non--Gaussianity tests described in Secs. \ref{sec:pairwise} and \ref{sec:ngchisq} to the above non-Gaussian summary statistics and show the results in the bottom rows of Figs. \ref{fig:ngmat} and \ref{fig:chisqs}. We find prominent non-Gaussian pairs in the case of the marked power spectrum on large scales, and on pairs involving large and small scales. The calculation of the mark assigned to every particle requires information from some large scale, described by the parameter $R$. \cite{philcox_2020} showed that this creates a coupling between large and small scales, that may be behind this phenomenon. On the other hand, the marked power spectrum seems to pass the $\chi^2$ distributional test (see Fig. \ref{fig:chisqs}).

For the bispectrum, we do not find as many highly non--Gaussian pairs as we do in $\rm{Mk}$ or the ${\rm WST}$. However, in this case the overall non--Gaussianity revealed by the $\chi^2$ distributional test is significant as seen in Figure \ref{fig:chisqs}. To check the robustness of our estimator for the bispectrum, we repeated the analysis from the public bispectrum measurements from the Quijote suite (Figure \ref{app:bk}), findind similar results. When using the $\chi^2$ test, we find substantial non Gaussianities for both bispectra measurements (see Fig. \ref{fig:chisqs}). We note that the presence of non-Gaussianities in the bispectrum likelihood was already noted in \cite{Scoccimarro_2000}.

For the WST, Figure \ref{fig:ngmat} reveals that several principal components have non Gaussian correlations with almost all other coefficients. At this point, it is hard to reveal whether these non--Gaussianities are caused by some small amount of coefficients or a combination of them since we apply a dimensionality reduction using the principal components (See Appendix \ref{app:WST}). However, a linear transformation of a (multivariate) Gaussian distributed variable is still Gaussian distributed, thus these non--Gaussianities should exist in the original coefficients. However, we warn the reader that since the amplitude of the z-scores is not dramatically large, these results may be affected by some inaccuracies as in the case of the power spectrum. Figure \ref{fig:chisqs} shows that the t values from the WST also deviates from the expected distribution in a manner similar to the bispectrum.

\subsection{Corrected Fisher Analysis}

As we did for the power spectrum, we \textit{Gaussianize} the above non-Gaussian statistics using the procedure described in Sec. \ref{sec:gaussianizing}. With the derived statistics, we perform a Fisher matrix analysis and show the results in the right panel of Figure \ref{fig:pkmkbkWST}. The results for the power spectrum are also plotted in Figure \ref{fig:pkmkbkWST} for reference. Table \ref{tab:change_table} contains the standard and corrected constraints while their ratio can be found in Table \ref{tab:change_r_table}.

The marked power spectrum's parameter constraints are affected by the correction. We find that the constraints on $\Omega_m$, $\Omega_b$, $\Omega_m$ and $\sigma_8$ has roughly doubled. It is also worth noting that the constraints from a $3\sigma$ Gaussian threshold to a $5\sigma$ condition are non negligible for the $\rm{Mk}$ as can be seen in Table \ref{tab:change_r_table}. We suspect that a large portion of the non--Gaussian components in Figure \ref{fig:pkmkbkWST} (a) are between these thresholds. It is interesting to see that the constraint on the neutrino mass is less affected then the other parameters and still is very promising compared to the power spectrum, at least for this analysis on the 3D matter density field.

For the the bispectrum, the parameter constraints are also affected resulting in constraints roughly $100\%$ bigger (less constraining) as we can see from Table \ref{tab:change_r_table}. The constraint on the neutrino mass ($M_\nu$), which is an important motivation for the bispectrum \citep{Hahn_2020}, is affected by $170\%$, making it only different by a $10\%$ level from the constraints from $\rm{log(Pk)}$. 
One could expect a similar effect for the halo or galaxy bispectra in redshift-space, see~\citep{Hahn_2020, Hahn_2021}. The extent to which this effect appears, however, would have to be estimated explicitly, and we make no claims about this in this work.

We originally had a intuition that the WST would have high levels of non--Gaussianity similarly to the bispectrum, since the same frequency components appear in the construction of several coefficients (see Appendix \ref{app:WST}). However, as we can read off Table \ref{tab:change_r_table}, the parameter change ratios were roughly similar to that of the power spectrum, except the case of the neutrino mass. It could be the case that our principal component selection actually removed most of the complex non--Gaussianities. Nevertheless, we find corrections roughly at the $50\%$ level, which cannot be overlooked.

We emphasize once again that the derived constraints from the Gaussianized statistics should be seen as a very conservative bound since the procedure we use to Gaussianize an statistic removes information. A full validation of the original constraints from the Fisher matrix would require to compare them against methods that do not throw away information.

\begin{deluxetable*}{l|llllllllllll}[h!]
\tablewidth{.9\textwidth}
\tablecaption{Standard and Corrected parameter constraints. For each parameter and statistic, we describe the standard and $3\sigma$ corrected marginalized fractional constraints. The $\delta$ are the percentage error on the reported ratios of changes when using different Gaussian mocks. The neutrino mass has a fiducial value of $M_\nu = 0$, so we write down the raw constraint. NA represents ``not applicable" and is the case where the Gaussianity test leaves less than 6 dimensions. \label{tab:change_table}}
\tablehead{&$\frac{\Delta \Omega_m}{\Omega_m}$&$\frac{\Delta \Omega_m}{\Omega_m}^{c}$&$\frac{\Delta \Omega_b}{\Omega_b}$&$\frac{\Delta \Omega_b}{\Omega_b}^{c}$&$\frac{\Delta h}{h}$&$\frac{\Delta h}{h}^{c}$&$\frac{\Delta n_s}{n_s}$&$\frac{\Delta n_s}{n_s}^{c}$&$\frac{\Delta \sigma_8}{\sigma_8}$&$\frac{\Delta \sigma_8}{\sigma_8}^{c}$&$\Delta M_{\nu}\:[eV]$&$\Delta M_{\nu}^{c}\:[eV]$}
\startdata
$\rm{Pk}$&0.271&0.433&0.752&1.109&0.683&1.038&0.463&0.73&0.014&0.023&0.789&1.273\\
$\delta$[\%]&&2.9&&1.4&&1.7&&1.7&&1.7&&1.0\\
\hline
$\rm{log(Pk)}$&0.273&0.319&0.758&0.851&0.689&0.784&0.467&0.54&0.014&0.016&0.786&0.912\\
$\delta$[\%]&&1.1&&0.7&&0.9&&1.0&&0.4&&0.5\\
\hline
$\rm{Pk}\oplus\rm{log(Pk)}$&0.057&0.35&0.245&0.934&0.19&0.853&0.094&0.589&0.003&0.018&0.082&0.987\\
$\delta$[\%]&&8.2&&6.6&&6.8&&7.2&&5.2&&5.9\\
\hline
$f(\rm{Pk})$&0.062&NA&0.216&NA&0.11&NA&0.055&NA&0.001&NA&0.082&NA\\
$\delta$[\%]&&NA&&NA&&NA&&NA&&NA&&NA\\
\hline
$\rm{Mk}$&0.042&0.083&0.212&0.39&0.147&0.275&0.05&0.101&0.002&0.005&0.017&0.025\\
$\delta$[\%]&&8.9&&7.8&&9.8&&4.3&&7.1&&3.5\\
\hline
$\rm{Bk}$&0.099&0.209&0.321&0.598&0.27&0.518&0.166&0.326&0.009&0.014&0.276&0.755\\
$\delta$[\%]&&3.3&&3.7&&3.5&&3.7&&10.2&&3.7\\
\hline
${\rm WST}$&0.087&0.129&0.404&0.591&0.259&0.372&0.056&0.086&0.002&0.003&0.058&0.111\\
$\delta$[\%]&&6.5&&2.8&&3.6&&2.5&&4.1&&10.4\\
\enddata
\end{deluxetable*}

\begin{deluxetable*}{l|llllllllllll}[h!]
\tablewidth{.9\textwidth}
\tablecaption{Parameter constraint change ratio. For each parameter and statistic,  we describe the ratio of the new constraint to the original constraints when applying a $3\sigma$ condition of non--Gaussianity and a $5\sigma$ condition. The $\delta$ are the percentage error on the reported ratios of changes when using different Gaussian mocks. Since the $5\sigma$ condition must reject less terms of a statistic, it is by construction more constraining than the $3\sigma$ condition while allowing more non--Gaussianity.  NA represents ``not applicable" and is the case where the Gaussianity test leaves less than 6 dimensions. \label{tab:change_r_table}}
\tablehead{&$\frac{\Delta \Omega_m^{c3}}{\Delta \Omega_m}$&$\frac{\Delta \Omega_m^{c5}}{\Delta \Omega_m}$&$\frac{\Delta \Omega_b^{c3}}{\Delta \Omega_b}$&$\frac{\Delta \Omega_b^{c5}}{\Delta \Omega_b}$&$\frac{\Delta h^{c3}}{\Delta h}$&$\frac{\Delta h^{c5}}{\Delta h}$&$\frac{\Delta n_s^{c3}}{\Delta n_s}$&$\frac{\Delta n_s^{c5}}{\Delta n_s}$&$\frac{\Delta \sigma_8^{c3}}{\Delta \sigma_8}$&$\frac{\Delta \sigma_8^{c5}}{\Delta \sigma_8}$&$\frac{\Delta M_\nu^{c3}}{\Delta M_\nu}$&$\frac{\Delta M_\nu^{c5}}{\Delta M_\nu}$}
\startdata
$\rm{Pk}$&1.596&1.518&1.474&1.427&1.52&1.47&1.576&1.509&1.623&1.518&1.615&1.516\\
$\delta$[\%]&2.9&1.6&1.4&1.0&1.7&1.1&1.7&0.8&1.7&3.6&1.0&2.7\\
\hline
$\rm{log(Pk)}$&1.165&1.137&1.122&1.101&1.137&1.113&1.156&1.13&1.16&1.145&1.16&1.143\\
$\delta$[\%]&1.1&0.4&0.7&0.4&0.9&0.4&1.0&0.4&0.4&0.3&0.5&0.3\\
\hline
$(Pk\oplus log(Pk))$&6.115&5.949&3.81&3.726&4.486&4.38&6.26&6.093&6.065&5.857&12.114&11.726\\
$\delta$[\%]&8.2&8.0&6.6&6.4&6.8&6.5&7.2&7.0&5.2&5.8&5.9&6.4\\
\hline
$f(\rm{Pk})$&NA&NA&NA&NA&NA&NA&NA&NA&NA&NA&NA&NA\\
$\delta$[\%]&NA&NA&NA&NA&NA&NA&NA&NA&NA&NA&NA&NA\\
\hline
$\rm{Mk}$&1.971&1.522&1.837&1.558&1.866&1.544&2.018&1.749&2.348&1.845&1.402&1.232\\
$\delta$[\%]&8.9&4.8&7.8&5.2&9.8&4.8&4.3&3.6&7.1&3.4&3.5&1.5\\
\hline
$\rm{Bk}$&2.103&1.638&1.864&1.464&1.921&1.509&1.957&1.538&1.678&1.341&2.736&2.111\\
$\delta$[\%]&3.3&1.8&3.7&0.8&3.5&1.1&3.7&1.2&10.2&1.4&3.7&1.4\\
\hline
$WST$&1.48&1.242&1.462&1.217&1.434&1.214&1.53&1.272&1.745&1.417&1.902&1.567\\
$\delta$[\%]&6.5&1.0&2.8&4.2&3.6&1.8&2.5&4.1&4.1&4.6&10.4&3.5\\
\enddata
\end{deluxetable*}

\section{Limitations of Gaussian tests}
\label{sec:hngc}

In this section we describe some of the limitations of the method and tests used to 1) identify non-Gaussianities, and 2) \textit{Gaussianize} the statistics.

In a case where one dimension is exactly a linear combination of other dimensions, the redundancy manifests in an obvious way (e.g. a large condition number or singular covariance). However, our example of $\rm{Pk} \oplus \rm{log(Pk)}$ is an instructive example of a non--Gaussian likelihood evading this check. The case here is more pernicious -- the information is redundant but in a non-linear way, which does not appear as an extremely large condition number. Nevertheless, the pairwise test makes it rather obvious which dimensions of the likelihood will cause the Gaussian approximation to break down. 
But, in the case of inputs derived from WST, a neural network, or some other complicated statistical probes, the issue is further complicated for two reasons
\begin{enumerate}
    \item The presence of non--linear relations between its dimensions cannot be easily guessed as it is the case for $\rm{Pk} \oplus \rm{log(Pk)}$, where we do suspect such a relation from construction.
    
    \item Such a relation could be a non--linear combination of many dimensions which can be hard to detect by the pairwise Gaussianity test.
\end{enumerate}

And thus, although our correction scheme renders the distribution of these statistic more compatible to a Gaussian approximation, we expect there to be many different ways a statistic could be non--Gaussian while evading the pairwise test.
As a simple example, we point out that a three component relation cannot be easily picked up with this test. Lets consider
\begin{align}\label{eq:ex}
    a&\sim\mathcal{N}(0,1)\\
    b&\sim\mathcal{N}(0,1)\\
    c&=\frac{a+b+\epsilon\times(a+b)^3}{\sqrt{2}}~.
\end{align}
When $\epsilon$ is zero, every 2 dimensional joint distribution will be a exactly a multivariate Gaussian while the full 3 dimensional distribution is clearly not. In fact, the covariance will be singular in this case, and this is something which can be easily spotted. However, when $\epsilon$ is not zero but small, the covariance will not be singular nor have a very big condition number. Every 2--dimensional sub distributions will still be very close to Gaussian and thus the pairwise non--Gaussianity test will fail to detect the severe non--Gaussianity. Extending the pairwise non--Gaussianity test to a triplet test would reveal the relation, however, this approach does not scale well with the dimensionality of the probe. 
Although this toy example seems to be artificially tailored to show this effect, similar cases are expected to show up in real data. The $\epsilon=0$ case is rarely seen in real data since such an explicit linear relation is usually discovered using linear analysis. However, nearly linear relations with slight non--linearities are expected to be a common case, even though the non--linear components might not be of any known form as the example above.
In general, for a d--dimensional statistical probe, if one can predict a single dimension of the statistic using the $d-1$ dimensions better than what a Gaussian process could do, a hidden relation between the dimensions of the statistics should be suspected to exist.

\begin{figure}[h!]
    \centering
    \includegraphics[width=70mm]{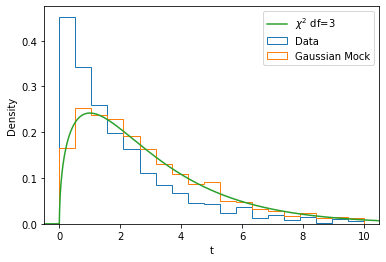}
    \caption{The non--Gaussianity of the joint distribution of a, b, c in Equation \ref{eq:ex} revealed by the $\chi^2$ distributional test.\label{fig:ex}}
\end{figure}

Further elaborating on this example, the non--Gaussianity here is detected by our $\chi^2$ distributional test, as we can see from Figure \ref{fig:ex}. This is because even though every 2--dimensional sub distribution is Gaussian, the $t$ values of the samples are not consistent with a Gaussian likelihood. Our $\chi^2$ distributional test thus serves as a complementary test to the pairwise Gaussianity test.

In the statistical probes explored, our $\chi^2$ distributional test was effective and indispensable in picking out the non--Gaussianity for $f(\rm{Pk})$, $\rm{Bk}$, and ${\rm WST}$, but we note that the test is somewhat less sensitive than the pairwise test. In complex statistics like the bispectrum and the WST, there could be very complicated hidden relations connecting some dimensions in a complicated highly non--Gaussian manner. In this sense, our $\chi^2$ distributional test is a good way to complement the pairwise test.

Finally, it should be noted that passing these tests should be treated as a necessary condition but not a sufficient one.
There are many ways non--Gaussianities can hide in high dimensional distributions, while we only check for the cases where 1) two dimensions of the statistic have a non--Gaussian partial distribution and 2) where the sharpness of the likelihood peak of the Gaussian approximation is vastly different from the one computed from data.

Lets consider the example of the  $f(\rm{Pk})$ statistic. While this statistic cannot contain more information than the power spectrum, the results from the Fisher may be interpreted in the other direction even if the Gaussian tests are passed. This clearly illustrate the limitations of the proposed tests.
In general, one should thus always simultaneously check for convergence, numerical stability and Gaussianity when performing a Fisher analysis and be as rigorous as possible.
We thus highlight that the interpretability of a statistical probe has a major importance, especially when using machine learning typed approaches, since they provide a intuition of how the joint distribution will behave.

\section{Conclusions}
\label{sec:conclusions}

The Fisher matrix formalism is commonly used in cosmology to quantify the accuracy that a given statistic can constrain the value of some cosmological parameters. This method will determine the variance of the optimal unbiased estimator for the considered statistic. However, the Fisher matrix is usually computed assuming that the statistic considered follows a multivariant Gaussian distribution. 

In this work we have considered several statistics to characterize the large-scale structure of the Universe and investigated whether their distribution is Gaussian or deviate from it. For this, we made use of two tests that will identify pairwise and global non-Gaussian distributions of the considered statistic. These tests can be employed in general and are not only designed for Fisher matrix calculations. We found non-Gaussianities in traditional statistics like the power spectrum and bispectrum but also in more recent statistics like the marked power spectrum and WST. We note that our conclusions are in agreement with previous works that have investigated this in depth \citep[see e.g.][]{Chang_2019}.

Next, we have applied a procedure to \textit{Gaussianize} the statistics, that consists in identifying the non-Gaussian components of the statistic and removing them. We stress that this procedure removes non-Gaussian dimensions, rather than Gaussianize the entire statistic. We have then performed Fisher matrix calculations with the standard and the Gaussianized statistics. We find significant corrections to the parameter constraints: $(62\%,\:51\%)$ for the power spectrum, $(134\%,\:84\%)$ for the marked power spectrum, $(173\%,\:111\%)$ for the bispectrum and  $(90\%,\:56\%)$ for the WST when the threshold to Gaussianize the statistics is set to $(5\sigma, 3\sigma)$, respectively.

We have also shown that without imposing Gaussinity for a given statistic, one can achieve unrealistically tight constraints on the value of the parameters. We illustrated this by considering the statistics $\rm{Pk} \oplus \rm{log(Pk)}$ (the concatenation of the power spectrum and the logarithm of it) and $f(\rm{Pk})$, that performed better than the power spectrum just by a non-linear transformation that do not contain cosmological information.

We have also outlined the limitations of the method we use in this work, that can identify pairwise and global (around the peak) non-Gaussianities, but that cannot identify more complex non-Gaussianities (e.g. higher-order interactions). It is also important to mention that we find that the Gaussianized statistics perform worse in constraining the value of the parameters. An obvious reason for this is because our method throws away the non-Gaussian information. A fairer comparison will be to develop an optimal method to Gaussianize a given statistic or to perform the inference with a method that do not rely on a Gaussian assumption, e.g. likelihood-free inference \cite[see e.g.][]{infomax_2018, Alsing_2019, Lucas_2021, Ana_2020}. Thus, the degraded constraints derived in this work from the Gaussianized statistics should be recalled as a conservative and perhaps more robust bound. This work however emphasizes the need to compare the constraints derived from the Fisher matrix with methods that do not discard the non-Gaussian information. 

We note that other methods may be more efficient at Gaussianize statistics. For instace, \cite{Scoccimarro_2000} proposed to use the PCA components of the bispectrum as a way to compress the relevant information and at the same time take advantage of the central limit theorem to Gaussianize the likelihood. We note that this strategy is similar to the one we have used for the WST, although the $\chi^2$ test revealed the presence of non-Gaussianities.

In general, Fisher matrix calculations are known to perform well at the 10\% level. In this work we have shown that under more conservative assumptions the Fisher constraints can be trusted within a factor of $\sim2$, at least for the statistics considered in this work. The tests used in this work can thus be used to quantify the robustness of the considered statistics to Fisher matrix assumptions. 

In the quest to find the best statistic to constraint the value of the cosmological parameters, it is important to keep mind the inherent limitations of the Fisher matrix formalism. The method used in this work will allow to complement the standard analysis with a more conservative Fisher matrix calculation. These, combined with methods like simulation based inference can help the community to identify robust statistics to retrieve the cosmological information from the large-scale structure of the Universe.

We note that the Gaussianity of a given statistic not only affects the outcome of Fisher matrix calculations, but traditional analyses performed using, for instance, Markov Chain Monte Carlo(MCMC) methods \cite[see e.g.][]{Philcox_2022,Byun_2021} commonly assume a Gaussian likelihood. If this assumption breaks down, corrections to the inferred parameters would also be expected.

We release the code we have used to compute the power spectra, bispectra, and WST. The code can be found in \url{https://github.com/cfpark00/LazyWaveletTransform} and works in both CPUs and GPUs.

\section*{Acknowledgements}
C.F.P acknowledges the support of NIH U01. Authors thanks Nayantara Mudur and Daniel Eisenstein for helpful discussions. Computations in this work was done on the Harvard FASRC Cannon Cluster.

\vspace{5mm}
\facilities{Harvard RC Cannon Cluster Computing}.

\software{NumPy \citep{harris2020array},
          SciPy \citep{2020SciPy-NMeth},
          PyTorch \citep{paszke2017automatic},
          SCIKIT--LEARN \citep{sklearn_api},
          H5PY \citep{collette_python_hdf5_2014},
          MATPLOTLIB \citep{Hunter:2007}
          }.

\appendix
\section{Parameters of the simulations}\label{app:sim_params}

Table \ref{tab:sims} contains the characteristics of the Quijote N-body simulations used for the Fisher matrix calculations in this work. We refer the reader to \cite{Villaescusa_Navarro_2020} for further details on the Quijote simulations.

\begin{deluxetable}{l|llllllllllll}[h!]
\tablewidth{.9\textwidth}
\tablecaption{The parameters of the simulations. N is the number of simulations and L denotes the size of the box in comoving units. IC denotes the method of initial condition generation. \label{tab:sims}}
\tablehead{Name&N&L[$h^{-1}\rm{Mpc}$]&IC&$\Omega_m$&$\Omega_b$&$h$&$n_s$&$\sigma_8$&$M_{\nu}[eV]$}
\startdata
Fiducial&N&1000&2LPT&0.3175&0.049&0.6711&0.9624&0.834&0.0\\
$\Omega_m^+$&15000&1000&2LPT&\textbf{0.3275}&0.049&0.6711&0.9624&0.834&0.0\\
$\Omega_m^-$&500&1000&2LPT&\textbf{0.3075}&0.049&0.6711&0.9624&0.834&0.0\\
$\Omega_b^+$&500&1000&2LPT&0.3175&\textbf{0.051}&0.6711&0.9624&0.834&0.0\\
$\Omega_b^-$&500&1000&2LPT&0.3175&\textbf{0.047}&0.6711&0.9624&0.834&0.0\\
$h^+$&500&1000&2LPT&0.3175&0.049&\textbf{0.6911}&0.9624&0.834&0.0\\
$h^-$&500&1000&2LPT&0.3175&0.049&\textbf{0.6511}&0.9624&0.834&0.0\\
$n_s^+$&500&1000&2LPT&0.3175&0.049&0.6711&\textbf{0.9824}&0.834&0.0\\
$n_s^-$&500&1000&2LPT&0.3175&0.049&0.6711&\textbf{0.9424}&0.834&0.0\\
$\sigma_8^+$&500&1000&2LPT&0.3175&0.049&0.6711&0.9624&\textbf{0.849}&0.0\\
$\sigma_8^-$&500&1000&2LPT&0.3175&0.049&0.6711&0.9624&\textbf{0.814}&0.0\\
$M_{\nu}$&500&1000&Zeldovich&0.3175&0.049&0.6711&0.9624&0.834&\textbf{0.0}\\
$M_{\nu}^{+}$&500&1000&Zeldovich&0.3175&0.049&0.6711&0.9624&0.834&\textbf{0.1}\\
$M_{\nu}^{++}$&500&1000&Zeldovich&0.3175&0.049&0.6711&0.9624&0.834&\textbf{0.2}\\
$M_{\nu}^{+++}$&500&1000&Zeldovich&0.3175&0.049&0.6711&0.9624&0.834&\textbf{0.4}\\
\enddata
\end{deluxetable}

\section{Details of the Statistics}\label{app:details}
\subsection{Power Spectrum}

We use the well know ``FFT and bin" method to compute the power spectrum. We bin the squared amplitudes into a uniform bins spaced by the frequency resolution: $k_{res}=2\pi/L$ where $L$ is the length of the box. For the sake of clarity, our bin edges are $[-0.5\:k_{res},\:,0.5\:k_{res},\:,\cdots,\left (\sqrt{3}\lfloor{H/2}\right \rfloor+0.5)\:k_{res}]$ for a grid side H. The factor of $\sqrt{3}$ comes from the 3D nature of the grid. Although, as it is clear from the above, we bin all the modes resulting from a FFT, to avoid contamination from any information from $|\vec{k}|>k_{Ny}$, we only use the bins below $0.5\: k_{Ny}$

\subsection{f(Pk)}\label{app:fpk}

We discuss the details of our information maximizing neural network. 
We use a multi layer perceptron architecture with a ReLU activation \citep{relu_citation}. A single sample of the input power spectrum with 78 elements, $\bm{Pk}\in\mathcal{R}^{78}$, is processed as following:
\begin{align*}
    \bm{x_1}&=(\bm{Pk}-\bm{\mu})\oslash \bm{\sigma}\\
    \bm{x_2}&=\rm{ReLU}(\bm{\Theta}_1 \bm{x_1}+\bm{b_1})\\
    \bm{x_3}&=\rm{ReLU}(\bm{\Theta}_2\bm{x_2}+\bm{b_2})\\
    \bm{x_4}&=\bm{\Theta}_3\bm{x_3}+\bm{b_3}\\
    f&:= \bm{Pk} \rightarrow \bm{x_4}
\end{align*}
where $\bm{x2},\bm{x3}\in\mathcal{R}^{32}$, $\bm{x4}\in\mathcal{R}^{20}$ and $\oslash$ is the Hadamard division. The matrices $\bm{\Theta}_i$ have dimensions compatible for the vector dimensions. The scaling vectors $\mu,\:\sigma$ are fixed to the fiducial mean and standard deviation. This transformation alone is a linear transform and thus does not affect the Fisher analysis up to numerical effects. However, neural networks perform optimally when the data is $\mathcal{O}(1)$ motivating this transform.\\
Calling the vector of all parameters, $\{\bm{\Theta}_1,\bm{\Theta}_2,\bm{\Theta}_3,b_1,b_2,b_3\}$ as $\bm{\lambda}$, we optimize for 
\begin{equation*}
    \mathcal{L}(\bm{\lambda})=\mathcal{L}_{\Delta\theta}(\bm{\lambda})+\mathcal{L}_{\rm{NG}}(\bm{\lambda})+\mathcal{L}_{\rm{Cond}}(\bm{\lambda})
\end{equation*}
using the Adam optimizer \citep{kingma2014adam} until convergence. $\mathcal{L}_{\Delta\theta}(\bm{\lambda})$ is simply defined as the sum of the marginalized parameter constraints, $\mathcal{L}_{NG}(\bm{\lambda})$ quantifies the dimension--wise non--Gaussianity of the samples, it is the squared sum of the skew and the kurtosis. $\mathcal{L}_{Cond}(\bm{\lambda})$ is the condition number of the covariance matrix times a small constant, here $0.001$. To avoid any potential issues from over fitting to these specific realizations of the simulation, we only use 70\% of the simulations.

\subsection{Marked Power Spectrum}

We use the mark in Equation \ref{eq:mark}, with optimal parameters $R = 10\:\rm{h^{-1}Mpc}$, $p = 2$, and $\delta_s = 0.25$. We do not calculate this but use the publicly available data from \cite{Villaescusa_Navarro_2020}.

\subsection{Bispectrum}\label{app:bk}

Here are our choice when implementing Equation \ref{eq:bk} with the FFT estimator \citep{sefusatti_2005,watkinson_2017}. 
\begin{itemize}
    \item Our computational representation of $\delta(|\vec{k}|-k))$, or a ``k--ring" centered at $k$. is smoothed. We use a b--splined kernel as the WST(see \ref{app:WST}) falling off to 0 at the center of the neighboring bins. It is normalized to unity.
    \item We use 16 k--values uniformly sampled in $\rm{log}(k)$ up to $k_{\rm{Ny}}$.
    \item We use all possible triangles satisfying the triangular inequality.
\end{itemize}
We end up with 825 valid configurations.

\subsection{Reduced Wavelet Scattering Transform}\label{app:WST}

We describe our 3D Wavelet scattering Transform(WST) and its reduction scheme. We start with wavelets similar to the Morlet wavelets in 2D. Then we multiply a 1D wavelet in the z-direction. Thus we treat the z--axis to be different than x and y motivated by the line of sight (LOS) direction in observational scenario. One could interpret these as 2.5D wavelets having one dimension different than the two others. For NZ LOS wavelets, NR radial wavelets and NT angular wavelets, our 3D wavelets are $\psi^z_{k}\times\psi^{xy}_{ij}$(x,y,z are just naming labels) where $0 \leq k < \rm{NZ}$, $0 \leq i < \rm{NR}$,$0 \leq j < \rm{NT}$, and we add the DC wavelets $\psi^{\rm{DC}}_k$ to keep the squared sum of the wavelets unity near the DC frequency.\\
Although one can use any LOS, radial, angular separations, we use radial bins and LOS bins equally spaced in logarithmic space and equally spaced angular bins:
\begin{align}
    r_i&=\begin{cases}
    0\:\rm{when}\: i < 0\\
    \rm{pow}(2,1+i*\frac{\rm{log}_2(k_{\rm{Ny}})-1}{\rm{NR}})\:\rm{when}\:0\leq i < \rm{NR}\\
    \rm{log}_2(k_{\rm Ny})\:\rm{when}\: \rm{NR} \leq i
    \end{cases}\\
    \theta_j&=j*\frac{\pi}{\rm NT}\\
    z_k&=\begin{cases}
    0\:\rm{when}\: i < 0\\
    \rm{pow}(2,1+i*\frac{\rm{log}_2(k_{\rm Ny})-1}{\rm NZ})\:\rm{when}\:0\leq i < \rm NZ\\
    \rm{log}_2(k_{\rm Ny})\:\rm{when}\: NZ \leq i
    \end{cases}\\
\end{align}
where $k_{Ny}$ is the Nyquist frequency. Note that we define the values at all i in order to simplify our wavelet definitions below\\
We then define the b--spline:
\begin{align}
b_0(t)&=\begin{cases}
1\:\rm{when}\: t< 0\\
2t^3-3t+1\:\rm{when}\: 0 \leq t < 1\\
0\:\rm{when}\: 1\leq t
\end{cases}\\
b_1(t,t_i,t_f)&=b_0(\frac{t-t_i}{t_f-t_i})\\
b_2(t,t_c,t_l,t_r)&=\begin{cases}
0\:\rm{when}\: t< t_l\\
b_1(t,t_c,t_l)\:\rm{when}\: t_l\leq t <t_c\\
b_1(t,t_c,t_r)\:\rm{when}\: t_c \leq t < t_r\\
0\:\rm{when}\: t_r\leq t\\
\end{cases}\\
\end{align}

Our wavelets are then:
\begin{align}
    \psi^{xy}_{ij}(r,\theta)&=b_2(r,r_i,r_{i-1},r_{i+1})\times b_2(\theta,\theta_i,\theta_{i-1},\theta_{i+1})\\
    \psi^{z}_{k}(z)&=b_2(z,z_i,z_{i-1},z_{i+1})\\
    \psi_{ijk}(r,\theta,z)&=\psi^{xy}_{ij}(r,\theta)\times\psi^{z}_{k}(z)
\end{align}
to where we add the DC wavelets
\begin{align}
    \psi^{DC}_k(r,\theta,z)=\begin{cases}
    0\:\rm{when}\:r_0 <r\\
    (1-\sum_{i,j}\psi^{xy}_{ij}(r,\theta))\times \psi^{z}_k(z)
    \end{cases}
\end{align}

We thus have $NF=(1+NR\times NT)\times NZ$ wavelets. 

Indicing all wavelets as $\bm{\psi}_i$(note that we abuse the notation to here represent the function sampled at all pixels needed to match the image size), the wavelet coefficients for an input image $\bm{I}$ are defined as:
\begin{align}
    \mu&=\rm{mean}(\bm{I})\\
    \sigma&=\rm{std}(\bm{I})\\
    \overrightarrow{S0}&=\{\mu,\sigma\}\\
    \bm{I^0}&=\frac{\bm{I}-\mu}{\sigma}\\
    \bm{I^1}_i&=\bm{I^0}\circledast\bm{\psi}_i\\
    \rm{S1}_i&=\rm{mean}(\bm{I^1}_i)\\
    \bm{I^2}_{ij}&=|\bm{I^0}_i|^2\circledast\bm{\psi}_j\\
    S2_{ij}&=\rm{mean}(\bm{I^2}_{ij})\\
    \overrightarrow{\rm{WST}}&=\{\overrightarrow{S0},\overrightarrow{\rm{S1}},\overrightarrow{\rm{S2}}\}
\end{align}

The mean and std operations are ran over the image domains, all bold fonts represent images, super scripted arrows represent vectors.

The total number of WST coefficients are then $2+\rm{NF}+\rm{NF}^2$. For this analysis, we discard the field mean which should be zero for all the overdensity fields. 
Due to the high dimensionality, we reduce the dimensionality by reporting the angular averaged coefficients. 
In detail, we report $S1$ coefficients averaged over angles. We then divide all $S2$ coefficients by the corresponding $S1$ coefficient to remove redundant information. We then only take the coefficients where the angular index and the LOS index for the first convolution and the second convolution are the same. We then take the averaged coefficient over this angle.
For a Fisher analysis, it would be useful to have an even smaller dimensionality due to numerical effects and convergence. We thus report only the first 500 principal components derived from the set of coefficients from the fiducial simulations. 
These coefficients have strictly less information than the whole coefficients. 
In practice, we deviate from the original wavelet scattering transform introduced by \citep{bruna2012invariant} in two senses:
\begin{enumerate}
    \item We use b--spline wavelets, these wavelets are indexed by $R$ and $T$ where $R$ represents the radial index and the $T$ represents the angular index.
    \item We take the absolute value squared instead of the absolute value as the non--linear operation between convolutions.
\end{enumerate}

We use b--spline wavelets for some motivations:
\begin{enumerate}
    \item $\sum_{ij} \psi_{ij}^2+\psi_{\rm{DC}}^2=1$ everywhere in the Fourier disk up to $|\vec{k}|=k_{max}$. The wavelets square sums to unity up to $k_{max}$. Thus the summed coefficients are expected to be more isotropic.  Then it decays to zero from $k_{max}$ to $k_{Ny}$.
    \item In addition to the above property, the wavelets decays to zero from $|\vec{k}|=k_{max}$ to $|\vec{k}|=k_{Ny}$, and thus we have no contribution at all from any modes over $k_{Ny}$.
    \item The wavelets decays precisely to zero within a sparse region of the Fourier space. This allows us much faster computation using the bounding boxes of the wavelets.
\end{enumerate}

\section{Computational Details}

We make our library computing Pk, Bk, WST\_abs, WST\_abs2 publicly available. Our code is available on \href{https://github.com/cfpark00/LazyWaveletTransform}{Github}. 

We discuss some details we consider to accelerate the computation.
\begin{enumerate}
    \item All our functions are batched. Modern machines' RAM and GPUs can easily have  $>10\:GB$ of memory. To perform FFTs and array slicing in an efficient manner, we batch every function. For a 2D field one thus feeds in a (C,H,W) array, and for a 3D field one feed in a (C,H,W,D) array.
    \item When the non--linearity applied between convolutions is the modulus squared for the WST, we do not perform the last IFFT since we can use the Plancherel theorem:
    \begin{equation*}
        \int_{-\infty}^{\infty} |f(x)|^2 dx=\int_{-\infty}^{\infty} |\hat{f}(k)|^2 dk
    \end{equation*}
    We can thus simply output the squared power multiplied by the wavelet in Fourier space.
    \item Most of our wavelets are sparse in Fourier Space. One can use a sparse representation and extract the relevant pixels of the Fourier space image. However, since sparse operations are inherently slower than dense operations, we use a much faster alternative. We exploit the fact that our wavelets are not only sparse in Fourier space but also compactly packed in a small region. (It is important that one computationally works in the Fourier space representation where the zero frequency is in the middle of an array.). We simply pre--compute the rectangular bounding box of each wavelet and only operate on the pixels in the bounding box. The remaining sparcity after extracting out only the bounding box is order 1.
    \item Since a Fourier transform of an image is Hermitian, we sample angles only up to $\pi$ and not $2\pi$. The coefficients will be the same anyways. In the 3D case, we choose to keep the angular sampling the same, and thus we cannot exploit the Hermitianity again in the z direction.
    \item For the power spectrum, we precompute the radial binning function and save it in memory as a sparse matrix. For a grid of side $H$, dimension $d$ and $NR$ radial bins, we have a matrix $P\in \mathcal{R}^{(H^d,NR)}$ where $P_{ij}$ is 1 if the $i^{th}$ pixel of the flattened the d--dimensional array falls into the $j^{th}$ bin. We also save the normalization needed, which doesn't depend on the input field.
\end{enumerate}

The runtime of our statistic code is described in Table \ref{tab:speed}.
\begin{deluxetable}{l|llllllllllll}
\tablewidth{.9\textwidth}
\tablecaption{Runtime of our statistics code on a NVIDIA A100-SXM4-40GB for double precision arithmetics\label{tab:speed}}
\tablehead{&\multicolumn{8}{c}{Time in ms}\\
&\multicolumn{4}{c}{2D}&\multicolumn{4}{c}{3D}\\
&\multicolumn{2}{c}{H=128}&\multicolumn{2}{c}{H=256}&\multicolumn{2}{c}{H=128}&\multicolumn{2}{c}{H=256}\\
&N=10&N=100&N=10&N=100&N=1&N=5&N=1&N=5}
\startdata
Pk&$0.5\pm 0.2$&$0.6\pm 0.2$&$0.6\pm 0.2$&$1.0\pm 0.5$&$2.7\pm 0.2$&$3.6\pm 0.2$&$18\pm 1$&$25\pm 1$\\
Bk (4096 configs)&$108\pm 6$&$285\pm 2$&$143\pm 5$&$1164\pm 8$&$390\pm 20$&$1860\pm 30$&$2900\pm 50$&$14100\pm 200$\\
WST(NR=4, NT=4)&$13\pm 2$&$14\pm 2$&$13\pm 2$&$30\pm 2$&-&-&-&-\\
WST(NR=8, NT=8)&$145\pm 6$&$146\pm 5$&$145\pm 4$&$168\pm 4$&-&-&-&-\\
WST(NR=4, NT=4, NZ=3)&-&-&-&-&$103\pm 7$&$176\pm 4$&$284\pm 5$&$1188.8\pm 0.1$\\
WST(NR=8, NT=8, NZ=6)&-&-&-&-&$5390\pm 60$&$5430\pm 20$&$5410\pm 30$&$12220\pm 20$\\
\enddata
\end{deluxetable}

\section{Cross check of our Fisher Matrices}

To confirm that we did not make a mistake in our analysis, we check our standard power spectrum Fisher matrix with the Fisher results published in \citep{Villaescusa_Navarro_2020}. We note that although very close, they are not numerically identical, we suspect that this results from the binning choice of the power spectrum.
\begin{figure*}[t!]
\gridline{\fig{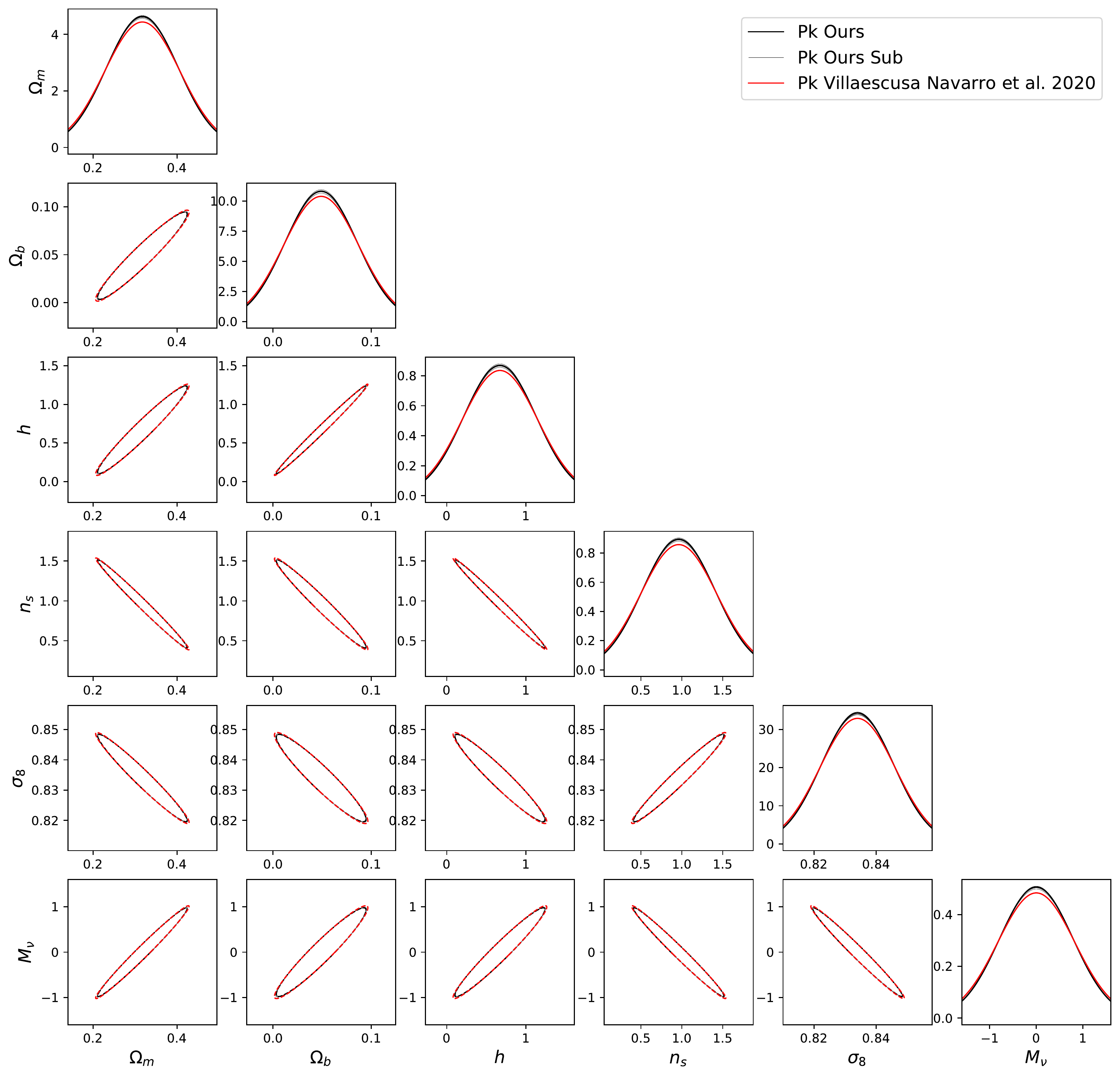}{.7\textwidth}{}}
\caption{Comparison of Fisher matrix constraints from our power spectrum and the power spectrum from \cite{Villaescusa_Navarro_2020}\label{fig:pk_comparison}}
\end{figure*}

\section{Check with an external bispectrum}

\begin{figure}[t!]
\gridline{\fig{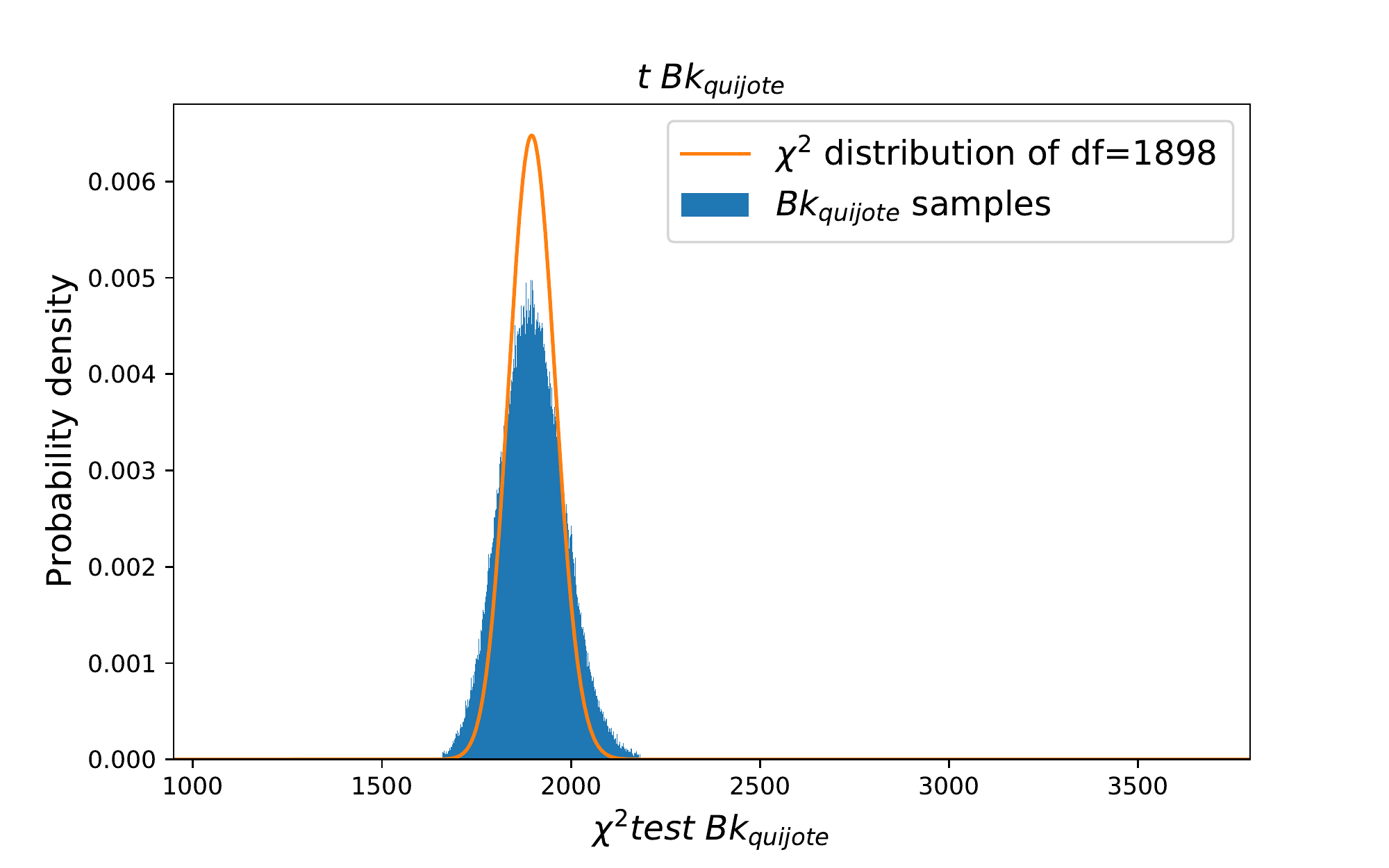}{.45\textwidth}{} }
\caption{$\chi^2$ distributional test of applied on the 1898 dimensional bispectrum from \cite{Villaescusa_Navarro_2020}\label{fig:chisqs_bk_quijote}}
\end{figure}

We do notice that small choices when implementing the FFT estimator \citep{sefusatti_2005,watkinson_2017} could alter the Fisher analysis results and could thus be important. Thus, it could be the case that only the bispectrum coefficients we obtained with \textit{our} code shows non--Gaussianity. If so, the $\chi^2$ distributional test would only be detecting non--Gaussianity from our own products. To verify that the finding generalizes to previously published bispectra, in this case from the Quijote suite products \cite{Villaescusa_Navarro_2020}, we apply the $\chi^2$ distributional test here. As we see in Figure \ref{fig:chisqs_bk_quijote}, we clearly detect the non--Gaussianity in this bispectrum.

\newpage

\bibliography{sample631}{}
\bibliographystyle{aasjournal}

\end{document}